\documentclass[aps,prd,twocolumn,10pt,preprintnumbers,amsmath,amssymb,
    superscriptaddress,floatfix,nofootinbib]{revtex4-2}

\makeatletter
\ifx\eqncolsep\undefined
  \newlength{\eqncolsep}
  \let\eqnarray\eqnarray@fleqn@fixed
\fi
\makeatother

\allowdisplaybreaks[4]

\usepackage{hyperref}
\hypersetup{
    colorlinks=true,
    linkcolor=blue,
    anchorcolor=blue,
    citecolor=blue,
    urlcolor=blue
}

\usepackage{physics}

\usepackage{graphicx}
\usepackage{booktabs}
\usepackage{array}
\usepackage{multirow}

\newcommand{\ucas}{\affiliation{School of Physical Sciences, University of
Chinese Academy of Sciences (UCAS), Beijing 100049, China}}

\newcommand{\cssm}{\affiliation{Special Research Centre for the Subatomic
Structure of Matter (CSSM), Department of Physics, Adelaide University,
Adelaide, 5005, South Australia, Australia}}

\begin{document}

\preprint{ADP-26-19/T1316}

\title{\textbf{Study of the $N^*(1535)$ and $N^*(1650)$ resonances from $\pi N$
$S_{11}$ scattering data and lattice QCD spectra}
}%

\author{Zhi-Xuan Li}
\ucas

\author{Derek B. Leinweber}\email{derek.leinweber@adelaide.edu.au}
\cssm

\author{Anthony W. Thomas}\email{anthony.thomas@adelaide.edu.au}
\cssm

\author{Jia-Jun Wu}%
\email[Corresponding author: ]{wujiajun@ucas.ac.cn}
\ucas

\date{November 20, 2026}

\begin{abstract}
We investigate two of the lowest-lying odd-parity nucleon resonances,
$N^*(1535)$ and $N^*(1650)$, within a nonperturbative Hamiltonian framework
(NPHF).
The coupled-channel model contains two three-quark bare states with total quark
spins $1/2$ and $3/2$, together with four meson-baryon channels: $\pi N$, $\eta
N$, $K\Lambda$, and $K\Sigma$.
Constraints from SU(3) flavor symmetry and the bare-state spin wave functions
are imposed on the bare-state--channel vertices and the channel-channel
Weinberg--Tomozawa interactions. This substantially reduces the number of free
parameters.
We combine fits to infinite-volume $\pi N$ $S_{11}$ scattering data with
finite-volume lattice QCD spectra. We use Hamiltonian eigenvectors to analyze
the model-basis compositions. A contamination function then compares the
resulting operator-mixing pattern with lattice correlation-matrix eigenvectors.
Within the scenarios considered, both bare states and coupled-channel
interactions play important roles in describing the two resonances.
From our combined infinite and finite volume analysis, in contrast to
conventional quark-model expectations, the spin-$1/2$ bare state is heavier than
the spin-$3/2$ bare state.
The contamination analysis is consistent with the associated operator assignment
and with a recent analysis of the relativistic wavefunctions associated with
these states.
The symmetry-constrained NPHF provides a unified description of the two $S_{11}$
resonances. The combined analysis also constrains their model interpretation.
\end{abstract}

\maketitle

\section{Introduction}
\label{intro}

Excited nucleons probe the nonperturbative regime of quantum chromodynamics
(QCD).
Their internal structure remains an open problem in hadron physics.
The two lowest odd-parity nucleon resonances, $N^*(1535)$ and $N^*(1650)$,
appear in the $\pi N$ $S_{11}$ partial wave. They remain longstanding puzzles in
light-hadron spectroscopy.
Their composition may involve three-quark components, dynamically generated
meson-baryon states, or mixtures of these components.
Different theoretical frameworks, from constituent quark models to
coupled-channel effective field theories and lattice QCD, yield distinct and
sometimes contradictory descriptions of these states~\cite{Kaiser:1995cy,
Nacher:1999vg, Bijker:2000gq, Inoue:2001ip, Zhao:2006an, Bruns:2010sv,
ParticleDataGroup:2026mpi, Stokes:2019zdd}.
A combined analysis with complementary constraints is therefore needed to
distinguish these interpretations.

Lattice QCD provides first-principles information on hadron structure and
complements experimental studies.
Although the ground-state octet and decuplet spectra are well established,
excited states remain more difficult to determine.
Lattice QCD calculations are performed in a finite volume, in Euclidean
spacetime, and often at unphysical quark masses.
Extracting physical information requires a relation between finite-volume
spectra and infinite-volume scattering amplitudes.
L\"uscher's formalism~\cite{Luscher:1990ux, Luscher:1986pf} provides such a
relation and is widely used to analyze lattice QCD spectra.

Scattering measurements and lattice QCD provide complementary constraints on
this system.
Infinite-volume $\pi N$ $S_{11}$ data densely sample the energy dependence of
the amplitude and constrain its phase shift and
inelasticity~\cite{Workman:2012hx}.
However, they do not directly determine the full set of coupled-channel
amplitudes that shape the resonance structure.
Finite-volume lattice QCD spectra encode coupled-channel effects but usually
contain only a few levels per ensemble and are often obtained at unphysical pion
masses.
A combined analysis can therefore constrain features that neither input
determines alone.

A method for combining infinite-volume scattering phenomenology with
finite-volume lattice QCD results was developed in Ref.~\cite{Abell:2023nex}
within the NPHF.
The NPHF gives a Hamiltonian formulation of the finite-volume problem that is
equivalent to L\"uscher's formalism for the systems considered
here~\cite{Wu:2014vma}.
The analysis of Ref.~\cite{Abell:2023nex} proceeds in three stages.
First, experimental scattering data at the physical pion mass constrain the
model parameters.
Second, a linear pion-mass dependence is introduced for the bare-state
parameters. The resulting finite-volume spectra are then compared with lattice
QCD results over a range of pion masses.
Third, the Hamiltonian eigenvectors are used to study the model-basis
composition of the finite-volume eigenstates. A contamination function then
compares the NPHF results with lattice correlation-matrix eigenvectors.
This procedure links scattering data to lattice spectra and provides
complementary constraints on resonance models.
Applications include the $N^*(1440)$ Roper resonance~\cite{Liu:2016uzk,
Wu:2017qve}, the $\Delta(1232)$ and $\Delta(1600)$~\cite{Hockley:2024ipz}, and
the $N^*(1535)$ and $N^*(1650)$~\cite{Abell:2023nex, Liu:2015ktc}. The method
has also been used for the $\Lambda(1405)$~\cite{Liu:2016wxq, Hall:2014uca} and
the charmed-strange mesons $D_{s0}^*(2317)$ and
$D_{s1}(2460)$~\cite{Yang:2021tvc}.

The model of Ref.~\cite{Abell:2023nex} follows a largely data-driven strategy
and does not fully impose symmetry constraints on the interaction vertices.
In particular, its bare states are not explicitly related to quark-model spin
configurations. The bare-state--channel coupling ratios and the channel-channel
Weinberg--Tomozawa (WT) interactions are also not constrained by SU(3) flavor
symmetry.
Consequently, the symmetry-motivated vanishing coupling between the spin-$3/2$
three-quark bare state and the $K\Lambda$ channel is not imposed.
These choices leave more free parameters.
The $K\Sigma$ channel is also absent from the coupled-channel basis, as it was
not easily constrained in their data-driven approach.
However, this setup limits the connection between the resonance structure and
the underlying quark-level symmetries.

In this work, we extend the NPHF analysis of the $N^*(1535)$ and $N^*(1650)$
resonances. We construct a symmetry-constrained model with two bare states and
four coupled channels: $\pi N$, $\eta N$, $K\Lambda$, and $K\Sigma$.
The bare states are identified with three-quark configurations of total quark
spin $1/2$ and $3/2$. SU(3) flavor symmetry and the spin wave functions
constrain their relative couplings, including the vanishing $K\Lambda$ coupling
of the spin-$3/2$ bare state.
These constraints substantially reduce the number of free parameters relative to
earlier data-driven implementations.
Following Ref.~\cite{Abell:2023nex}, we first fit the $\pi N$ $S_{11}$
scattering data. We then extend the model to finite volumes and unphysical pion
masses and compare the spectra with lattice QCD results. Finally, we analyze the
eigenstate compositions and scattering-state contamination.
Together, these steps constrain the roles of the three-quark bare states and
meson-baryon channels in the two resonances.

The paper is organized as follows.
Section~II introduces the NPHF formalism, including the infinite-volume
scattering equation, the finite-volume Hamiltonian, and the symmetry-constrained
interactions.
Section~III presents fits to the infinite-volume $\pi N$ $S_{11}$ amplitude and
compares the model scenarios.
Finite-volume spectra and their comparison with lattice QCD results are
discussed in Sec.~IV.
Section~V presents the scattering-state contamination analysis.
Section~VI summarizes our conclusions and outlines further tests.

\section{Formalism}

\subsection{Hamiltonian model}
In the center-of-mass frame, the Hamiltonian for the interacting meson-baryon
system is written as
\begin{eqnarray}
  \hat{H}=\hat{H}_0+\hat{H}_I ,
\end{eqnarray}
where $\hat{H}_0$ is the free Hamiltonian and $\hat{H}_I$ contains the
interactions.
The free Hamiltonian is
\begin{eqnarray}
        \label{noninh}
  \hat{H}_0&=&\sum_{i=1}^{n}\ket{N_i}m^0_{N_i}\bra{N_i} +\sum_{\alpha}\int
  d^3\vec{k} \nonumber\\
  &\;&\ket{\alpha(\vec{k})}\{
  \omega_{\alpha_M}(|\vec{k}|)+\omega_{\alpha_B}(|\vec{k}|) \}
  \bra{\alpha(\vec{k})} ,
\end{eqnarray}
where $N_i$ is the $i$th bare state with mass $m^0_{N_i}$. The index $\alpha$
labels the two-particle channels $\pi N$, $\eta N$, $K\Lambda$, and $K\Sigma$.
The meson or baryon energies in channel $\alpha$ are
$\omega_{\alpha_{M(B)}}(|\vec{k}|)=
\sqrt{m_{\alpha_{M(B)}}^2+\vec{k}^2}$, where $m_{\alpha_{M(B)}}$ denotes the
corresponding mass.

\begin{figure}[htbp]
    \centering
    \begin{minipage}[b]{0.2\textwidth}
        \centering
        \includegraphics[width=\linewidth]{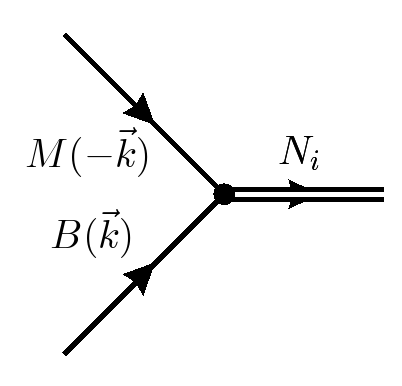}
        \par (a)
    \end{minipage}
    \hspace{.4in}
    \begin{minipage}[b]{0.2\textwidth}
        \centering
        \includegraphics[width=\linewidth]{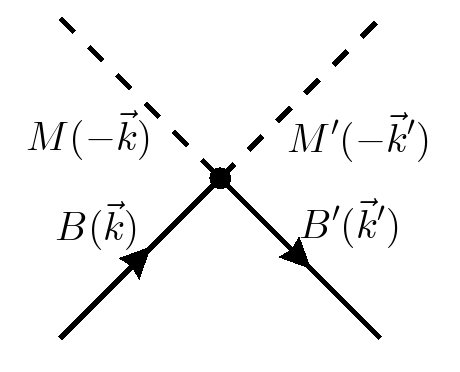}
        \par (b)
    \end{minipage}
    \caption{Diagrammatic representation of the bare-state--channel vertex
    $G_{\alpha}^{N_i}(k)$ (left) and the channel-channel interaction
    $V_{\alpha\beta}(k,k')$ (right) in the center-of-mass frame.}
    \label{fig:vertex}
\end{figure}

The interaction Hamiltonian contains the two contributions shown in
Fig.~\ref{fig:vertex}.
It reads
\begin{eqnarray}
  \label{eq:hefthi}
  \hat{H}_I=\hat{g}+\hat{v}.
\end{eqnarray}
Here, $\hat{g}$ couples the bare states to the two-particle channels:
\begin{equation}
  \label{eq:heftg}
    \begin{split}
        \hat{g}=\sum_{\alpha , i} \int d^3  \vec{k} \big\{ &
        \ket{N_i}\bar{G}_{\alpha}^{N_i}(\vec{k})\, \bra{\alpha(\vec{k})\,} +
        \mathrm{H.c.}\big\}\,,
    \end{split}
\end{equation}
where $\bar{G}_{\alpha}^{N_i}$ is the momentum-dependent vertex function.
The second term, $\hat{v}$, couples the two-particle basis states in channels
$\alpha$ and $\beta$:
\begin{eqnarray}
  \label{eq:heftv}
  \hat{v}=\sum_{\alpha,\beta} \int d^3 \vec{k}\, d^3 \vec{k}^{\prime}\,
  \ket{\alpha(\vec{k})\,} \bar{V}_{\alpha , \beta}(\vec{k},\vec{k}^\prime)\,
  \bra{\beta(\vec{k}^\prime)\,},
\end{eqnarray}
where $\bar{V}_{\alpha\beta}$ is the momentum-dependent channel-channel
interaction.
We take the channel structure of $V_{\alpha\beta}$ from the leading-order
interaction in chiral perturbation theory ($\chi$PT).

\subsection{Infinite-volume formalism}

The coupled-channel two-body $T$ matrix is obtained from the Lippmann--Schwinger
equation, a three-dimensional reduction of the Bethe--Salpeter equation.
After partial-wave projection, the equation is
\begin{eqnarray}
  \label{eq:BSeq}
  T_{\alpha\beta}(k,k^\prime;E)&=&\mathcal{V}_{\alpha\beta}(k,k^\prime;E)+%
  \sum_{\gamma}\int q^2\, dq\, \mathcal{V}_{\alpha\gamma}(k,q;E)\nonumber\\
  &\;&\times \frac{1}{E-\omega_\gamma (q)+i\varepsilon}\,
  T_{\gamma\beta}(q,k^\prime;E)\, ,
\end{eqnarray}
where $\omega_\gamma(q)=\omega_{\gamma_M}(q)+\omega_{\gamma_B}(q)$ is the energy
of a two-particle state in channel $\gamma$ with relative momentum $q$.
The interaction Hamiltonian gives the coupled-channel potential

\begin{equation}
\label{eq:tildeV}
    \begin{split}
        \mathcal{V}_{\alpha\beta}(k,k^\prime;E)=& \sum_{i}
        \frac{{G_{\alpha}^{N_i}}^\dagger(k)\,
        G^{N_i}_{\beta}(k^\prime)}{E-m^0_{N_i}}
        + V_{\alpha\beta}(k,k^\prime)\, ,
    \end{split}
\end{equation}
where $G$ and $V$ are the partial-wave projections of $\bar{G}$ and $\bar{V}$ in
Eqs.~(\ref{eq:heftg},\ref{eq:heftv}), respectively.
Because only $s$-wave contributions are retained, the partial-wave projected
potential differs from the full potential only by a factor of $4\pi$.
With the normalization
$\braket{\alpha(\vec{k})}{\beta(\vec{k}')}=\delta_{\alpha,\beta}\delta(\vec{k}%
-\vec{k}')$, the $S$ matrix is related to the $T$ matrix by
\begin{equation}
      \label{eq:Smatrix}
  S_{\alpha\beta}(E)=\delta_{\alpha\beta}-2\pi i\rho^{1/2}_{\alpha}(E)\,
  T_{\alpha\beta}(k_{\alpha}^{\mathrm{on}},k_{\beta}^{\mathrm{on}};E)\,
  \rho^{1/2}_{\beta}(E)\, ,
\end{equation}
where $k_{\alpha}^{\mathrm{on}}(E)$ is the on-shell momentum in channel
$\alpha$. For an open channel, the density of states is

\begin{eqnarray}
  \label{eq:densitystate}
  \rho_{\alpha}(E)=\frac{\omega_{\alpha_M}(k_{\alpha}^{\mathrm{on}})\,
  \omega_{\alpha_B}(k_{\alpha}^{\mathrm{on}})}{E} k_{\alpha}^{\mathrm{on}} ,
\end{eqnarray}

The inelasticity $\eta_\alpha$ and phase shift $\delta_\alpha$ are defined
through
\begin{eqnarray}
  \label{eq:inelasity}
  S_{\alpha\alpha}(E)= \eta_\alpha(E)\, e^{2i\delta_\alpha(E)} ,
\end{eqnarray}

Resonance poles are located by continuing the amplitude into the complex energy
plane.
Each channel has a branch point at its two-particle threshold and an associated
right-hand cut. Crossing the cut connects two Riemann sheets.
For $n$ coupled channels, the \(2^n\) sheets are labeled by an ordered sign
vector \((\pm,\pm,\dots,\pm)\). Here the channel order is \((\pi N,\eta
N,K\Lambda,K\Sigma)\). A plus sign (+) denotes the physical branch of the
channel momentum, while a minus sign (-) denotes its continuation across the
corresponding cut.
Finite-width resonance poles generally lie on sheets that are unphysical for at
least one open channel.
Let \(A_{N_i,N_j}^{(\text{RS})}\) denote the dressed bare-state propagator
matrix on a specified sheet. The pole energy is a zero of its inverse:
\begin{equation}
  \label{eq:pole-condition}
  \det\!\left[\left(A^{(\text{RS})}(E_\text{pole})\right)^{-1}\right]=0.
\end{equation}

\subsection{Finite-volume Hamiltonian}

For a cubic box of length $L$ with periodic boundary conditions, the momentum is
quantized as
\begin{eqnarray}
  \vec{k}_n=\frac{2\pi}{L}\vec{n} , \vec{n}=(n_x,n_y,n_z)\, ,
\end{eqnarray}
where $\vec{n}$ is a three-dimensional integer vector with
$n_x,n_y,n_z\in\mathbb{Z}$.
The infinite-volume integral is then replaced by
\begin{eqnarray}
  \int d^3 k\, f(\vec{k})\rightarrow
  \left(\frac{2\pi}{L}\right)^3\sum_{\vec{n}\in \mathbb{Z}^3}
  f\left(\frac{2\pi\vec{n}}{L}\right).
\end{eqnarray}

Within the $s$-wave truncation used here, momentum modes with the same value of
$n=\vec n^{\,2}$ can be grouped together.
The three-dimensional sum therefore reduces to
\begin{eqnarray}
  \sum_{\vec{n}\in \mathbb{Z}^3} \rightarrow \sum_{n \in \mathbb{N}_0} C_3 ( n
  )\, ,
\end{eqnarray}
where $C_3(n)$ is the number of integer vectors with squared norm
$n=n_x^2+n_y^2+n_z^2$, and the scalar shell momentum is $k_n=2\pi\sqrt{n}/L$.
For example, $C_3(0)=1$, $C_3(2)=12$, and $C_3(7)=0$.

We use finite-range regularization (FRR) to regulate the integrals and sums.
The regulator $u(k)$ is inserted into both the bare-state vertex and the
two-particle interaction to suppress high-momentum contributions.
The next subsection specifies the vertex and two-particle interactions.
The finite-volume Hamiltonian is formally infinite-dimensional.
Because \(u(k)\) suppresses high momenta, the matrix can be truncated in
numerical calculations.
The maximum momentum retained in the Hamiltonian matrix is denoted
\(k_{n_\text{max}}\).
We determine this cutoff by requiring the eigenvalues to be numerically
converged.
Specifically, \(k_{n_\text{max}}\) is fixed by the condition
\(u(k_{n_\text{max}})=u_\text{min}\).
We take \(u_\text{min}=10^{-2}\).
For a dipole form factor, the corresponding maximum momentum is
\(k_{n_\text{max}}=\Lambda\sqrt{u^{-1/2}-1}\),
with \(u=10^{-2}\).

We denote the bare model-basis states by $\ket{B_i}$. In the two-bare-state
scenarios, $(\ket{B_1},\ket{B_2})=(\ket{N_2},\ket{N_4})$. The matrices below
display this two-state case. A one-bare-state scenario follows by deleting the
$B_2$ row and column, while Scenario I contains only the meson-baryon block. In
this basis, the free Hamiltonian in Eq.~(\ref{noninh}) becomes

\begin{equation}
  \begin{split}
     H_0&= \operatorname{diag} \big\{ m_{B_1}^0 , m_{B_2}^0 , \omega_{\pi
     N}(0)\, , \omega_{\eta N}(0)\, , \omega_{K \Lambda}(0)\, ,  \\
   & \omega_{K \Sigma}(0)\, , \omega_{\pi N}(k_1)\, , \omega_{\eta N}(k_1)\, ,
   \omega_{K \Lambda}(k_1)\, , \omega_{K \Sigma}(k_1)\, , \cdots, \\
   & \omega_{\pi N}(k_{n_{\max}})\, , \omega_{\eta N}(k_{n_{\max}})\, ,
   \omega_{K \Lambda}(k_{n_{\max}})\, , \omega_{K \Sigma}(k_{n_{\max}})\,
   \big\}.
  \end{split}
  % Keep the tag below the last row with the affected amsmath releases.
  \IfPackageAtLeastTF{amsmath}{2024/03/11}{%
    \IfPackageAtLeastTF{amsmath}{2025/05/18}{}{\raisetag{-16.87004pt}}%
  }{}
\end{equation}

The interactions $G$ and $V$ in Eq.~(\ref{eq:tildeV}) become $\tilde{G}$ and
$\tilde{V}$ in finite volume.
The corresponding relations are~\cite{Wu:2014vma}
\begin{eqnarray}
  \tilde{G}_{\alpha}^{B_i} (k_n) &=& \sqrt{\frac{C_3 (n)}{4\pi}}\, \left(
  \frac{2\pi}{L} \right)^{\frac{3}{2}}\, G_{\alpha}^{B_i} (k_n)\, , \nonumber \\
  \tilde{V}_{\alpha\beta}(k_n,k_{n^\prime}) &=& \frac{\sqrt{C_3 (n)\, C_3
  (n^\prime)}}{4\pi}\, \left( \frac{2\pi}{L} \right)^{3}\,
  V_{\alpha\beta}(k_n,k_{n^\prime})\, ,\nonumber\\
\end{eqnarray}
where the factor $4\pi$ arises from the partial-wave projection of the pure
$s$-wave interaction.

The finite-volume interaction Hamiltonian can then be written as
\begin{widetext}
\begin{eqnarray}
  H_I=\begin{pmatrix}
    0 &0 &\tilde{G}_{\pi N}^{B_1}(0)  &\tilde{G}_{\eta N}^{B_1}(0) &\tilde{G}_{K
    \Lambda}^{B_1}(0) &\tilde{G}_{K \Sigma}^{B_1}(0) &\tilde{G}_{\pi
    N}^{B_1}(k_1) &\cdots  \\
    0 &0 &\tilde{G}_{\pi N}^{B_2}(0)  &\tilde{G}_{\eta N}^{B_2}(0) &\tilde{G}_{K
    \Lambda}^{B_2}(0) &\tilde{G}_{K \Sigma}^{B_2}(0) &\tilde{G}_{\pi
    N}^{B_2}(k_1) &\cdots  \\
    \tilde{G}_{\pi N}^{B_1}(0) &\tilde{G}_{\pi N}^{B_2}(0) &\tilde{V}_{\pi N,\pi
    N}(0,0) &\tilde{V}_{\pi N,\eta N}(0,0) &\tilde{V}_{\pi N,K \Lambda}(0,0)
    &\tilde{V}_{\pi N,K \Sigma}(0,0) &\tilde{V}_{\pi N,\pi N}(0,k_1) &\cdots \\
    \tilde{G}_{\eta N}^{B_1}(0) &\tilde{G}_{\eta N}^{B_2}(0) &\tilde{V}_{\eta
    N,\pi N}(0,0) &\tilde{V}_{\eta N,\eta N}(0,0) &\tilde{V}_{\eta N,K
    \Lambda}(0,0) &\tilde{V}_{\eta N,K \Sigma}(0,0) &\tilde{V}_{\eta N,\pi
    N}(0,k_1) &\cdots\\
    \tilde{G}_{K \Lambda}^{B_1}(0) &\tilde{G}_{K \Lambda}^{B_2}(0) &\tilde{V}_{K
    \Lambda,\pi N}(0,0) &\tilde{V}_{K \Lambda,\eta N}(0,0) &\tilde{V}_{K
    \Lambda,K \Lambda}(0,0) &\tilde{V}_{K \Lambda,K \Sigma}(0,0) &\tilde{V}_{K
    \Lambda,\pi N}(0,k_1) &\cdots\\
    \tilde{G}_{K \Sigma}^{B_1}(0) &\tilde{G}_{K \Sigma}^{B_2}(0) &\tilde{V}_{K
    \Sigma,\pi N}(0,0) &\tilde{V}_{K \Sigma,\eta N}(0,0) &\tilde{V}_{K \Sigma,K
    \Lambda}(0,0) &\tilde{V}_{K \Sigma,K \Sigma}(0,0) &\tilde{V}_{K \Sigma,\pi
    N}(0,k_1) &\cdots\\
    \tilde{G}_{\pi N}^{B_1}(k_1) &\tilde{G}_{\pi N}^{B_2}(k_1) &\tilde{V}_{\pi
    N,\pi N}(k_1,0) &\tilde{V}_{\pi N,\eta N}(k_1,0) &\tilde{V}_{\pi N,K
    \Lambda}(k_1,0) &\tilde{V}_{\pi N,K \Sigma}(k_1,0) &\tilde{V}_{\pi N,\pi
    N}(k_1,k_1) &\cdots\\
    \vdots   &\vdots  &\vdots &\vdots &\vdots &\vdots &\vdots &\ddots
  \end{pmatrix} .
\end{eqnarray}
\end{widetext}
All couplings are real in the present parametrization, so the displayed matrix
is symmetric.
The eigenvalues of $H=H_0+H_I$ give the discrete finite-volume energy levels and
satisfy $\det(H-E_i\mathbb{I})=0$.
The associated eigenvectors give the overlaps $\braket{B_j}{E_i}$. Their squared
magnitudes quantify the bare-state components of $\ket{E_i}$ within the chosen
model basis.

\subsection{Interaction vertex}

The direct two-particle interaction $V_{\alpha\beta}(k,k')$ is based on the
leading-order Weinberg--Tomozawa term of chiral perturbation
theory~\cite{Veit:1984sf}:
\begin{equation}
  \label{WT-int}
  \begin{aligned}
  V_{\alpha \beta} (k,k^\prime)
  ={}&-C_{\alpha \beta} \frac{g_0}{8\pi^2 f_\alpha f_\beta}
  u_\alpha(k)\, u_\beta(k^\prime) \\
  &\times\frac{\omega_{\alpha_M} (k)+\omega_{\beta_M} (k^\prime)}
  {\sqrt{2\omega_{\alpha_M}(k)}\, \sqrt{2\omega_{\beta_M} (k^\prime)}}\, .
  \end{aligned}
\end{equation}
where $f_\alpha$ is the decay constant of the meson in channel $\alpha$. We take
$f_\pi=92.4$~MeV, $f_K=1.2f_\pi$, and $f_\eta=1.3f_\pi$~\cite{Inoue:2001ip}.
The SU(3) Clebsch--Gordan coefficients $C_{\alpha\beta}$ are listed in
Table~\ref{tab:CCcouplingCG}.
The overall channel-channel coupling $g_0$ is fitted to the data.
We use the on-mass-shell prescription when solving the Lippmann--Schwinger
equation~\cite{Li:2024zld}. The factor
$\omega_{\alpha_M}(k)+\omega_{\beta_M}(k^\prime)$ in the WT vertex is evaluated
explicitly at the integration or summation momenta.

\begin{table}[h!]
\caption{SU(3) Clebsch--Gordan coefficients $C_{\alpha\beta}$ for the $s$-wave
meson-baryon interaction in the $I=1/2$ sector~\cite{Lu:2024ajt}.}
\label{tab:CCcouplingCG}
\begin{ruledtabular}
\begin{tabular}{ccccc}
$C_{\alpha\beta}$ & $\pi N$    & $\eta N$ & $K  \Lambda$  & $K \Sigma$    \\
\colrule
$\pi N$  & $2$ & $0$   & $-\frac{3}{2}$ & $-\frac{1}{2}$ \\
$\eta N$  & $0$  & $0$ & $-\frac{3}{2}$ & $\frac{3}{2}$ \\
$K  \Lambda$  & $-\frac{3}{2}$  & $-\frac{3}{2}$ & $0$ & $0$ \\
$K \Sigma$  & $-\frac{1}{2}$ & $\frac{3}{2}$   & $0$ & $2$ \\
\end{tabular}
\end{ruledtabular}
\end{table}

We regulate the ultraviolet behavior by inserting the FRR function $u(k)$ in
Eq.~(\ref{WT-int}).
Within the power-counting regime, FRR is equivalent to dimensionally regulated
chiral perturbation theory after the low-energy coefficients are
matched~\cite{Young:2002ib}. It also provides a finite spatial profile for the
source.
We choose a dipole regulator for each channel:
\begin{eqnarray}
  \label{dipole}
  u_{\alpha} (k) = (1+k^2/\Lambda_{\alpha}^2)^{-2} ,
\end{eqnarray}
where $\Lambda_{\alpha}$ is the regulator scale for channel $\alpha$.
After fitting the model to the experimental data, the remaining parameters
depend implicitly on the regulator scales $\Lambda_{\alpha}$.
Their fitted values can therefore change with the regularization scheme. In
particular, $g_0$ depends on $\Lambda_{\alpha}$ and need not remain fixed at
$1$.
This momentum-dependent potential is pseudo-separable: it is a finite sum of
separable terms. The Lippmann--Schwinger equation can therefore be solved
analytically in terms of a finite matrix of loop integrals.

The interaction between bare state $N_i$ and two-particle channel $\alpha$ is
described by a standard $s$-wave parametrization.
Motivated by heavy-baryon $\chi$PT, we write the vertex as~\cite{Wu:2014vma,
Liu:2015ktc}
\begin{eqnarray}
  \label{G-int}
  G_{\alpha}^{N_i} (k) = A_{\alpha}^{N_i} \frac{\sqrt{3} g_{N_i}}{2\pi f_\alpha}
  \sqrt{\omega_{M} (k)}\, u_{N_i} (k)\, ,
\end{eqnarray}
where $g_{N_i}$ is the overall coupling between bare state $N_i$ and the
pseudoscalar-octet meson-baryon channels, and $\omega_M$ is the energy of meson.
The coefficient $A_{\alpha}^{N_i}$ fixes the relative coupling to channel
$\alpha$.
If only SU(3) flavor symmetry is imposed, these coefficients are the same for
the $N^*(1535)$ and $N^*(1650)$.
The $8\to8\otimes8$ transition contains the symmetric coupling $D$ and the
antisymmetric coupling $F$.
Their relative strength is parametrized by $\alpha=F/(D+F)$.
We take the leading-order quark-model value
$\alpha=0.4$~\cite{Dashen:1994qi,Jiang:2022gjy}.%
The resulting channel coefficients are
\begin{equation}
  \label{GcouplingCGF}
  A_{\alpha}^{N_i} = (1, 1/5, -3/5, -1/5).
\end{equation}

SU(3) flavor symmetry alone does not distinguish the two excited nucleon states.
For the lowest $J^P=\frac{1}{2}^-$ excitations, the three-quark orbital angular
momentum is a $P$ wave, while the total quark spin can be $1/2$ or $3/2$.
The spin wave functions can therefore produce different meson-baryon couplings
for the two $N^*$ states.
At leading order, the total quark spin-$3/2$ state does not couple to the
$K\Lambda$ channel~\cite{Bijker:2000gq,Zhao:2006an}.
This illustrates the role of the spin structure in the coupling pattern.
To include this dependence, we construct the full wave functions of the two bare
states.
We denote them by $\ket{N_{2s+1}}$, where $s$ is the total quark spin.
Their wave functions are
\begin{widetext}
\begin{equation}
  \label{N7028}
  \ket{N_2}=\sum_{L_z+S_z =J_z} \braket{1L_z,\frac{1}{2}S_z}{
  \frac{1}{2}J_z}\times \frac{1}{2}[(\phi^\rho\, \chi_{S_z}^\lambda +
  \phi^\lambda\, \chi_{S_z}^\rho)\, \Psi_{11L_z}^\rho + (\phi^\rho\,
  \chi_{S_z}^\rho - \phi^\lambda\, \chi_{S_z}^\lambda)\, \Psi_{11L_z}^\lambda]\,
  ,
\end{equation}
\begin{equation}
    \label{N7048}
  \ket{N_4}=\sum_{L_z+S_z =J_z} \braket{1L_z,\frac{3}{2}S_z }{
  \frac{1}{2}J_z}\times \frac{1}{\sqrt{2}}[ \phi^\rho\, \chi_{S_z}^s\,
  \Psi_{11L_z}^\rho + \phi^\lambda\, \chi_{S_z}^s\, \Psi_{11L_z}^\lambda ] ,
\end{equation}
\end{widetext}
where $\phi$, $\chi$, and $\Psi$ denote the flavor, spin, and spatial wave
functions, respectively.
The labels $\rho$ and $\lambda$ denote components that are antisymmetric and
symmetric, respectively, under interchange of the first two quarks. The
superscript $s$ denotes a fully symmetric component.
The wave function of a ground-state octet baryon $\ket{B}$ is
\begin{equation}
  \label{N5628}
  \ket{B}=\frac{1}{\sqrt{2}}(\phi^\rho\, \chi^\rho + \phi^\lambda\,
  \chi^\lambda)\, \Psi_{000}^s\, .
\end{equation}

The coefficients $A_{\alpha}^{N_i}$ follow from matrix elements of the
flavor-spin operator between the initial and final states.
In the chiral quark model, the meson-quark coupling is described by an
axial-current interaction. The coefficient $A_{\alpha}^{N_i}$ is proportional to
the factor $\zeta_m$ defined in Ref.~\cite{Bijker:2000gq}:
\begin{equation}
 A_{\alpha}^{N_i} \propto \zeta_m = \sum\limits_\gamma  {\left\langle
 {\begin{array}{*{20}{c}}
{({p_f},{q_f})}&{(p,q)}\\
{{I_f},{Y_f}}&{I,Y}
\end{array}} \right|\left. {\begin{array}{*{20}{c}}
{{{({p_i},{q_i})}_\gamma }}\\
{{I_i},{Y_i}}
\end{array}} \right\rangle {\alpha _{m,\gamma }}},\label{chqm}
\end{equation}
where $\gamma$ labels the multiplicity of the representation $(p,q)$.
In the present case, it distinguishes the components of the baryon wave
function.
The SU(3) isoscalar factor in Eq.~(\ref{chqm}) depends on the flavor multiplet
$(p,q)$ and on the isospin $I$ and hypercharge $Y$ of the initial and final
baryons. It also depends on the emitted meson.
The factor $\alpha_{m,\gamma}$ contains the spin and angular-momentum dependence
and is given in Ref.~\cite{Bijker:2000gq}.

Evaluating Eq.~(\ref{chqm}) for transitions from the two bare states to the
meson-baryon channels gives the coefficients listed in
Table~\ref{tab:Gcoupling}.
The quark-pair-creation model gives a similar coupling pattern, indicating that
these ratios are not specific to one decay model.
In particular, $A_{K\Lambda}^{N_4}=0$, consistent with the symmetry argument
above~\cite{Zhao:2006an}.

\begin{table}[h!]
  \caption{Bare-state--channel coefficients $A_\alpha^{N_i}$ for the two
  quark-spin configurations.}
  \label{tab:Gcoupling}
  \begin{ruledtabular}
\begin{tabular}{ccccc}
$A_\alpha^{N_i}$ & $\pi N$    & $\eta N$ & $K  \Lambda$  & $K \Sigma$    \\
\colrule
$N_2$  & $\frac{\sqrt{2/3}}{3}$ & $\frac{1}{3\sqrt{6}}$   &
$-\frac{1}{2\sqrt{6}}$ & $\frac{1}{6\sqrt{6}}$ \\
$N_4$  & $\frac{1}{3\sqrt{6}}$  & $-\frac{1}{3\sqrt{6}}$ & $0$ &
$-\frac{\sqrt{2/3}}{3}$ \\
\end{tabular}
\end{ruledtabular}
\end{table}

For both interactions, $V$ and $G$, the overall couplings and regulator scales
are fitted. Their relative channel strengths are fixed by $C_{\alpha\beta}$ and
$A_{\alpha}^{N_i}$.
This substantially reduces the number of free parameters.
The full model has one coupling and four regulator scales for $V$. The two bare
states add two couplings, two bare masses, and two regulator scales for $G$.

\begin{table*}[t]
  \caption{
    Fit parameters, $\chi^2/\mathrm{dof}$, and pole positions for all scenarios.
    Scenario I contains four coupled channels and no bare states.
    Its $\chi^2/\mathrm{dof}$ is too large to be listed.
    Scenario II contains two bare states but no direct channel-channel
    interaction.
    Scenario III contains one bare state and four coupled channels.
    In Scenario III, case (a) uses $A^{N_i}_\alpha$ from
    Eq.~\eqref{GcouplingCGF}; cases (b) and (c) use the first and second rows of
    Table~\ref{tab:Gcoupling}, respectively.
    Scenario IV contains two bare states and four coupled channels.
    Case (A) uses the fixed quark-model mixing $\xi=-0.53$, whereas $\xi=0$ in
    cases (B) and (C).
    Because of the mixing, the asterisk-prefixed entries in the $m^0_{N_2}$ and
    $m^0_{N_4}$ rows denote the fitted physical-point bare-mass eigenvalues
    $m_{B_1}$ and $m_{B_2}$, respectively.
    Cases (B) and (C) denote the two retained unmixed local solutions.
    The three pole rows correspond, from top to bottom, to the Riemann sheets
    \texttt{-+++}, \texttt{--++}, and \texttt{---+}, with channel order $(\pi
    N,\eta N,K\Lambda,K\Sigma)$.
    Underlined entries lie at a fit boundary and are shown without errors.
  }
  \label{tab:scenarios-combined}

  \begingroup
  \setlength{\tabcolsep}{3pt}
  \renewcommand{\arraystretch}{1.08}
  \resizebox{\textwidth}{!}{%
  \begin{tabular}{@{}l*{8}{c}@{}}
    \toprule
    \multirow{2}{*}{\textbf{Parameter}}
      & \multirow{2}{*}{\textbf{I}}
      & \multirow{2}{*}{\textbf{II}}
      & \multicolumn{3}{c}{\textbf{III}}
      & \multicolumn{3}{c}{\textbf{IV}} \\
    \cmidrule(lr){4-6}
    \cmidrule(lr){7-9}
      & & &
      \textbf{a} & \textbf{b} & \textbf{c}
      & \textbf{A} & \textbf{B} & \textbf{C} \\
    \midrule
    $g_0$ & $0.519\pm0.0034$ & - & $0.713\pm0.0028$ & $0.695\pm0.0047$ &
    $1.664\pm0.0082$ & $0.713\pm0.0070$ & $0.711\pm0.0067$ & $0.711\pm0.0060$ \\
    $\Lambda_{\pi N}/\mathrm{GeV}$ & $0.574\pm0.0062$ & - & $0.404\pm0.0048$ &
    $0.445\pm0.0038$ & $0.433\pm0.0017$ & $0.435\pm0.0041$ & $0.419\pm0.0044$ &
    $0.444\pm0.0038$ \\
    $\Lambda_{\eta N}/\mathrm{GeV}$ & $\underline{1.600}$ & - &
    $\underline{1.600}$ & $1.596\pm0.070$ & $0.819\pm0.0090$ & $0.998\pm0.066$ &
    $1.480\pm0.046$ & $0.431\pm0.019$ \\
    $\Lambda_{K\Lambda}/\mathrm{GeV}$ & $\underline{1.600}$ & - &
    $0.723\pm0.036$ & $\underline{0.400}$ & $0.777\pm0.020$ &
    $\underline{0.400}$ & $\underline{0.400}$ & $\underline{0.400}$ \\
    $\Lambda_{K\Sigma}/\mathrm{GeV}$ & $\underline{1.600}$ & - &
    $\underline{1.600}$ & $\underline{1.600}$ & $\underline{1.600}$ &
    $1.376\pm0.046$ & $1.466\pm0.024$ & $1.249\pm0.023$ \\
    $m^0_{N_2}/\mathrm{GeV}$ & - & $1.649\pm0.0022$ & $1.762\pm0.012$ &
    $1.710\pm0.0026$ & - & $*\,\underline{1.750}$ & $\underline{1.750}$ &
    $1.905\pm0.016$ \\
    $m^0_{N_4}/\mathrm{GeV}$ & - & $1.993\pm0.0092$ & - & - &
    $\underline{1.400}$ & $*\,1.864\pm0.013$ & $1.866\pm0.034$ &
    $\underline{1.750}$ \\
    $g_{N_2}$ & - & $3.018\pm0.011$ & $0.265\pm0.0090$ & $0.913\pm0.013$ & - &
    $0.837\pm0.016$ & $0.972\pm0.017$ & $0.626\pm0.010$ \\
    $g_{N_4}$ & - & $0.930\pm0.013$ & - & - & $\underline{4.000}$ &
    $0.532\pm0.071$ & $0.318\pm0.032$ & $0.929\pm0.037$ \\
    $\Lambda_{N_2}/\mathrm{GeV}$ & - & $\underline{0.400}$ & $1.053\pm0.053$ &
    $0.930\pm0.015$ & - & $1.243\pm0.034$ & $0.991\pm0.015$ &
    $\underline{1.600}$ \\
    $\Lambda_{N_4}/\mathrm{GeV}$ & - & $\underline{1.600}$ & - & - &
    $0.637\pm0.0021$ & $1.375\pm0.14$ & $\underline{1.600}$ & $1.269\pm0.049$ \\
    \midrule
    $\chi^2/\mathrm{dof}$ & - & 67.948 & 8.148 & 16.811 & 44.776 & 6.133 & 6.785
    & 5.491 \\
    \midrule
    \multirow{3}{*}{Pole$/\mathrm{GeV}$} & - & $1.075-0.195\,\mathrm{i}$ &
    $0.997-0.153\,\mathrm{i}$ & $1.003-0.170\,\mathrm{i}$ &
    $1.032-0.145\,\mathrm{i}$ & $1.003-0.167\,\mathrm{i}$ &
    $1.000-0.162\,\mathrm{i}$ & $1.004-0.170\,\mathrm{i}$ \\
    \cline{3-9}
     & - & - & $1.466-0.100\,\mathrm{i}$ & $1.438-0.068\,\mathrm{i}$ & - &
     $1.497-0.066\,\mathrm{i}$ & $1.476-0.086\,\mathrm{i}$ &
     $1.529-0.055\,\mathrm{i}$ \\
    \cline{3-9}
     & - & $1.674-0.037\,\mathrm{i}$ & $1.696-0.039\,\mathrm{i}$ &
     $1.700-0.033\,\mathrm{i}$ & $1.708-0.050\,\mathrm{i}$ &
     $1.687-0.061\,\mathrm{i}$ & $1.684-0.048\,\mathrm{i}$ &
     $1.680-0.091\,\mathrm{i}$ \\
    \bottomrule
  \end{tabular}%
  }
  \endgroup
\end{table*}

\section{Infinite-volume fits to $\pi N$ $S_{11}$ phase shifts and
inelasticities}
\label{sec:infinite_volume_fits}

We first constrain the model parameters through global fits to the $\pi N$
$S_{11}$ partial-wave amplitude.

The present work concentrates on two nucleon resonances, $N^*(1535)$ and
$N^*(1650)$.
In the conventional quark model, they are associated with the two lowest
$P$-wave configurations, denoted by $\ket{N_2}$ and $\ket{N_4}$ in the preceding
section.
The two bare configurations are embedded in the $\pi N$, $\eta N$, $K\Lambda$,
and $K\Sigma$ coupled-channel system. Their vertices obey the symmetry relations
derived above, including the vanishing $\ket{N_4}$--$K\Lambda$ coupling. We test
whether the two resonances can be described by the interplay of the bare states
and these meson-baryon channels.
Experimental decay data show that the $N^*(1535)$ couples strongly to $\eta
N$~\cite{ParticleDataGroup:2026mpi}.
Phenomenological analyses of its couplings also suggest a large $s\bar{s}$
component~\cite{Liu:2005pm}, as well as various relevant reaction
analyses~\cite{Xie:2017erh,Xie:2007qt,Nieves:2001wt,Doring:2008sv,%
Pavao:2018wdf}.
These findings motivate an explicit treatment of coupled-channel effects.

Moreover, chiral effective field theory studies dynamically generate an
$N^*(1535)$ pole through coupled-channel rescattering, without an explicit bare
quark state~\cite{Inoue:2001ip, Bruns:2010sv}.
Motivated by these findings, we construct four scenarios within the formalism of
Sec.~II. We fit the $\pi N$ $S_{11}$ partial-wave amplitude to test which
physical interpretations are consistent with experimental data.

The four scenarios are as follows:
I. The four meson-baryon channels are included without a bare state;
II. Both bare states are included without direct channel-channel interactions;
III. One bare state and all four channels are included. The bare state is
assigned either no explicit spin structure, \(|N_2\rangle\), or \(|N_4\rangle\);
IV. Both bare states and all four channels are included, with different choices
of the mixing parameter $\xi$ and bare-mass ordering.

We fit the $\pi N$ $S_{11}$ partial-wave amplitude data from the SAID analysis
over 1100--1750~MeV~\cite{Workman:2012hx}.
The fit contains 40 points for each of its real and imaginary parts.
We determine the model parameters by minimizing
$\chi^2=\chi^2_{\mathrm{Re}T}+\chi^2_{\mathrm{Im}T}$, where the two terms are
the contributions from the real and imaginary parts of the $T$ matrix.
The resulting phase shift and inelasticity are then compared with the SAID
values.
For the retained fits, we search the relevant unphysical sheets in the lower
half-plane and compare the extracted pole positions with the Particle Data Group
(PDG) values.

Table~\ref{tab:scenarios-combined} lists the best-fit parameters, reduced
$\chi^2$ values, and pole positions for all scenarios.
The fits in Scenarios III(a) and IV, which contain both bare-state contributions
and direct channel-channel interactions, give the much lower
$\chi^2/\mathrm{dof}$ value.
This improvement also accompanies additional parameter freedom.
The scattering fit alone therefore does not identify the preferred dynamics.
In particular, Scenarios I and II deviate substantially from the data.

\begin{figure*}[ht]
  \centering
  \includegraphics[width=0.98\textwidth]{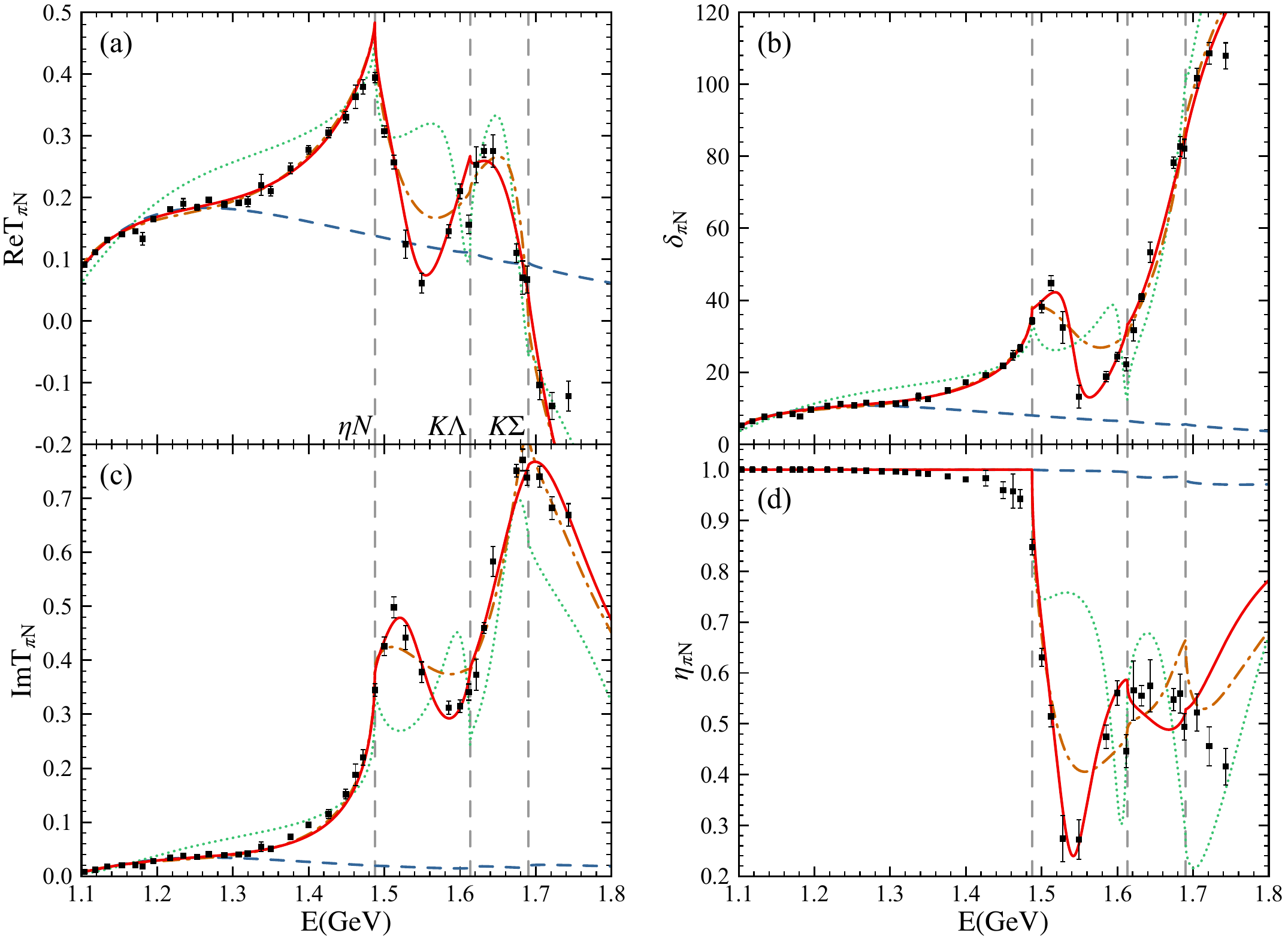}
  \caption{Real part of the $T$ matrix (a), phase shift (b), imaginary part of
  the $T$ matrix (c), and inelasticity (d).
  The black squares with error bars are from the SAID
  analysis~\cite{Workman:2012hx}.
  The three gray dashed lines mark the $\eta N$, $K\Lambda$, and $K\Sigma$
  thresholds.
  The red solid curves show the best fit for Scenario IV(C).
  The blue dashed curves show Scenario I.
  The green dotted curves show Scenario II.
  The brown dash-dotted curves show Scenario III(a).}
  \label{fig:sc1a2a3}
\end{figure*}

In Scenario I, the channel-channel interaction reproduces the data near the $\pi
N$ threshold, as shown by the blue dashed curves in Fig.~\ref{fig:sc1a2a3}. This
is the region where the leading chiral interaction is expected to be most
reliable.
By contrast, the bare-state $s$-channel exchanges in Scenario II reproduce the
broad shape of the $T$ matrix but not its detailed peaks and dips.
Thus, the bare-state exchanges account for the broad resonance structure, while
channel-channel interactions are needed for its detailed energy dependence.

Scenario III contains one explicit bare state. The second resonance pole is
obtained without a second explicit bare state and is generated by the
meson--baryon dynamics in this scenario.
We consider three bare-state assignments.
In case (a), the spin structure is omitted and $A_{\alpha}^{N_i}$ is given by
the SU(3) coefficients in Eq.~\eqref{GcouplingCGF}.
In case (b), the bare state is identified with $\ket{N_2}$, and
$A_{\alpha}^{N_i}$ is taken from the first row of Table~\ref{tab:Gcoupling}.
Case (c) instead uses the $\ket{N_4}$ coefficients in the second row.
Cases (b) and (c) retain only one of the two spin configurations.
They give much poorer fits than Scenario III(a), so we do not discuss them
further. The Scenario III(a) result is shown by the brown dash-dotted curves in
Fig.~\ref{fig:sc1a2a3}.

Although this single-bare-state model improves on Scenario II, two clear
discrepancies remain.
Between 1.5 and 1.6~GeV, the real and imaginary parts of the measured $T$ matrix
contain a pronounced dip. Scenario III(a) reproduces this structure
qualitatively but underestimates its depth.
Between 1.6 and 1.7~GeV, the peak in the real part also differs from the data.
Correspondingly, the phase shift is poorly described from 1.5 to 1.6~GeV, and
the inelasticity deviates from the data from 1.6 to 1.7~GeV.
Adding the second bare state substantially improves all of these features.

\begin{figure*}[ht]
  \centering
 \includegraphics[width=0.98\textwidth]{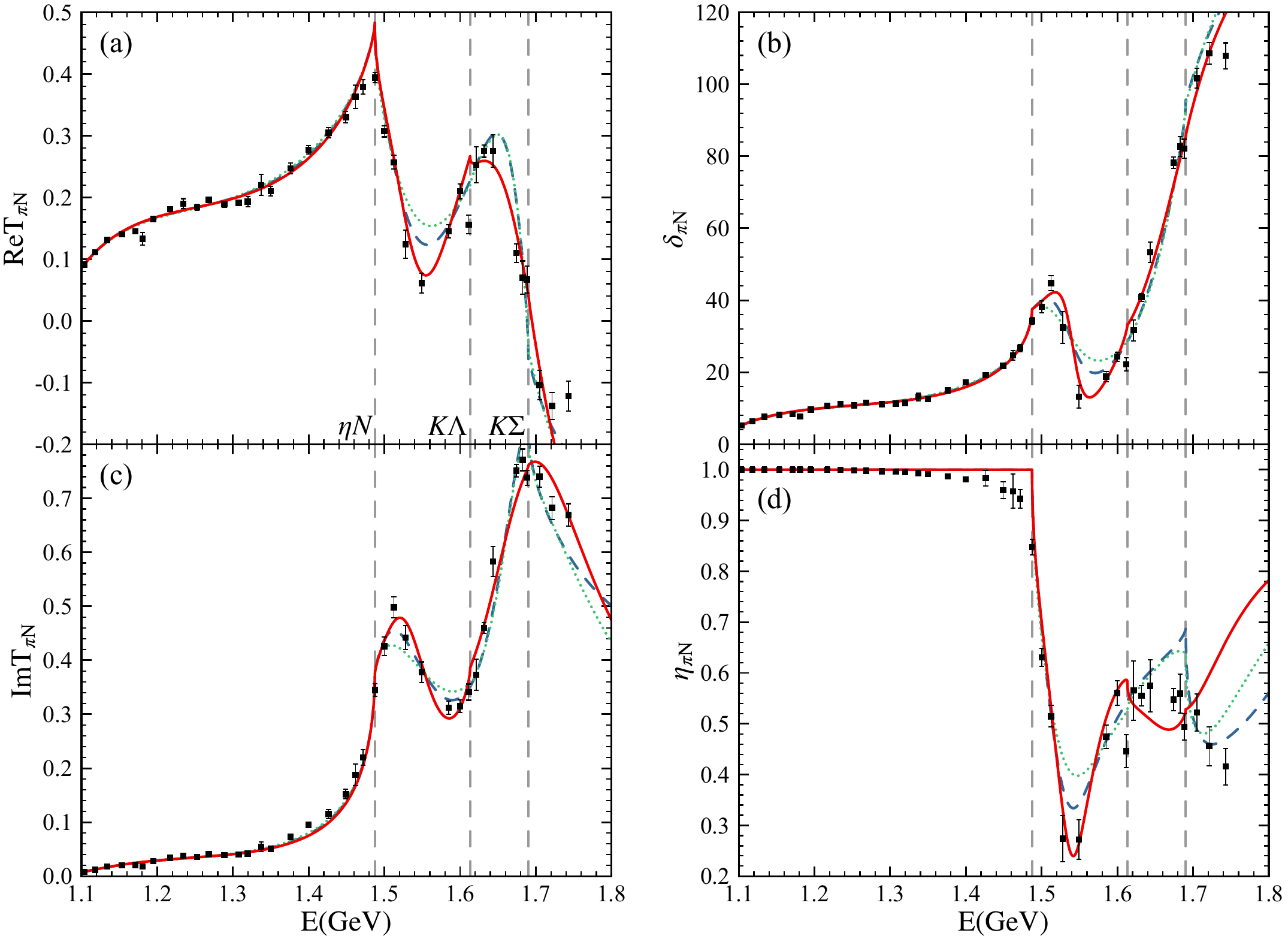}
  \caption{
  Real part of the $T$ matrix (a), phase shift (b), imaginary part of the $T$
  matrix (c), and inelasticity (d).
  The blue dashed curves show Scenario IV(A).
  The green dotted curves show Scenario IV(B).
  The remaining notation follows Fig.~\ref{fig:sc1a2a3}.}
  \label{fig:sc4}
\end{figure*}

Scenario IV contains both bare states and all four coupled channels.
The states $\ket{N_2}$ and $\ket{N_4}$ have the same conserved quantum numbers
and can therefore mix.
In the conventional quark model, this mixing is described by an angle
$\theta_S$.
We define $\xi=\sin\theta_S$, with $\sqrt{1-\xi^2}=\cos\theta_S$ in our
convention.
Diagonalizing the bare-state mass matrix yields the mixed bare-mass eigenstates
$\ket{B_1}$ and $\ket{B_2}$ with eigenvalues $m_{B_1}$ and $m_{B_2}$.
For nonzero $\xi$, the channel couplings are linear combinations of those for
the two quark-model configurations:
\begin{equation}
  \label{mixing}
  \begin{aligned}
    {g_{{B_1},\alpha }} &= \sqrt {1 - {\xi ^2}}\, {A^{N_2}_\alpha}\, {g_{N_2}} +
    \xi\, {A^{N_4}_\alpha}\, {g_{N_4}},\\
    {g_{{B_2},\alpha }} &=  - \xi\, {A^{N_2}_\alpha}\, {g_{N_2}} + \sqrt {1 -
    {\xi ^2}}\, {A^{N_4}_\alpha}\, {g_{N_4}},
  \end{aligned}
\end{equation}
where $A^{N_i}_{\alpha}$ are the symmetry coefficients in
Table~\ref{tab:Gcoupling}, and $g_{N_i}$ are the unmixed overall couplings
defined in Eq.~\eqref{G-int}.

Here we introduce three variants within Scenario IV.
First, adopting the quark-model result $|\theta_S|\simeq32^\circ$ (corresponding
to $|\xi|\simeq0.53$) from Ref.~\cite{Saghai:2009zza} gives Scenario IV(A).
Since that quark model fitted physical nucleon resonance states without
coupled-channel contributions, the mixing may stem from hadronic coupled-channel
effects.
Thus, at the quark-gluon level we may set $\xi=0$, which yields a fit with a
smaller $\chi^2/\text{dof}$ but reverses the bare-state mass ordering, i.e.,
$m^0_{N_2}>m^0_{N_4}$; we label this Scenario IV(C).
In the conventional quark model, one expects $m^0_{N_2}<m^0_{N_4}$; imposing
this constraint recovers a similar solution, which we call Scenario IV(B).

As shown in Table~\ref{tab:scenarios-combined} and Fig.~\ref{fig:sc4}, all three
subcases of Scenario IV yield much smaller $\chi^2/\mathrm{dof}$ values
($6.133$, $6.785$, and $5.491$ for cases (A), (B), and (C), respectively) and
closer agreement with data than the preceding scenarios, with a consistent
channel-channel coupling $g_0\approx0.71$.
The $T$-matrix curves of Scenarios IV(A) and IV(B) are nearly identical
(Fig.~\ref{fig:sc4}).
A limitation of cases (A) and (B) is that the dips in $\mathrm{Re}\,T_{\pi N}$
and $\delta_{\pi N}$ between $1.52$ and $1.60$ GeV are not well described,
whereas Scenario IV(C) shows significant improvement.
However, at energies above the $K\Sigma$ threshold, Scenario IV(C) fails to
describe the inelasticity well, suggesting an important role for additional
single- and two-particle channels at these higher energies.
Overall, Scenario IV(C) achieves the lowest $\chi^2/\mathrm{dof}=5.491$.

The solutions of Ref.~\cite{GWDataAnalysisCenter} are obtained from
global partial-wave analyses.
We use them here as benchmarks for the energy dependence of the $\pi N$
amplitude.
Evaluated on the amplitude points considered here, the benchmark solutions give
$\chi^2$ values ranging from 208 to 293, whereas Scenario~IV(C) gives $\chi^2
\simeq 379$ for 69 degrees of freedom.
Relative to the benchmark solution with $\chi^2 \simeq 208$, just four points
account for essentially the entire net difference of 171. Roughly 88 comes from
the point at the $\eta N$ threshold, which may reflect the omission of the
$\pi\pi N$ channel. The two points near the $K\Lambda$ threshold contribute
around 38, while 45 comes from the highest-energy point.
Near the $K\Lambda$ threshold, our curve follows one of the two points closely
but misses the other, while the benchmark curve passes between them.
Finally, at the top range of energies considered here one may expect new
contributions, including the effect of higher mass resonances.
Thus, Scenario~IV(C) provides a satisfactory description of the broad energy
dependence of scattering data in the $S_{11}$ channel.

We now turn to the fitted bare masses.
In Table~\ref{tab:scenarios-combined}, the fitted bare masses in Scenarios
IV(A)\textendash{}IV(C) lie in the $1.75$\textendash{}$1.91~\mathrm{GeV}$ range,
with the lighter bare state at $1.750~\mathrm{GeV}$.
In practice, the data-driven fit consistently demands a heavy mass for the
lower-mass state, sometimes pushing it directly to $2~\mathrm{GeV}$; we
therefore imposed an upper limit of $1.750~\mathrm{GeV}$.
In any case, these undressed bare masses are heavier than typical constituent
quark-model estimates for the lowest $P$-wave nucleon excitations
($\sim1.50$\textendash{}$1.65~\mathrm{GeV}$)~\cite{Isgur:1978xj,
Capstick:1986ter, Bijker:2000gq, Zhao:2006an}.
Within the NPHF, this upward shift is a direct consequence of coupled-channel
dressing: dispersive self-energy loops from the $\pi N$, $\eta N$, $K\Lambda$,
and $K\Sigma$ channels generate substantial negative mass shifts, requiring
heavier bare masses to place the dressed resonance poles near $1.535$ and
$1.650~\mathrm{GeV}$.
We also performed an additional test by fixing the lower bare mass to
$1.6~\mathrm{GeV}$ to match constituent quark-model expectations.
As demonstrated in Appendix~\ref{app:lower_bound}, the infinite-volume fits for
Scenarios IV(A) and IV(B) deteriorate sharply ($\chi^2/\mathrm{dof}=17.555$ and
$29.487$), whereas Scenario IV(C) retains an accurate description of the
scattering data ($\chi^2/\mathrm{dof}=6.908$).
These tests show that current data prefer Scenario IV(C), i.e.,
$m^0_{N_2}>m^0_{N_4}$, even when a lower bare mass is imposed.

In quark models, ground-state baryon octets with total spin $1/2$ are lighter
than spin-$3/2$ decuplets.
This splitting arises from the color-magnetic spin-spin interaction, which is
attractive for $S=1/2$ and repulsive for $S=3/2$; when this interaction
dominates, higher-spin baryons are heavier.
For the negative-parity resonances considered here, spin-orbit and tensor
interactions also contribute and both can lower the mass of $|N_4\rangle$,
counteracting the spin-spin splitting.
In standard one-gluon-exchange models, these corrections are typically too weak
to reverse the ordering.

On the other hand, relativistic nucleon wave functions from lattice studies in
Ref.~\cite{Mickley:2026zsi} support this inverted hierarchy.
With the operator assignments discussed in
Sec.~\ref{sec:scattering_contaminationformalism}, the state associated with
$|N_2\rangle$ has an additional node in its lower $s$-wave Dirac component,
whereas the state associated with $|N_4\rangle$ does not.
For $|N_2\rangle$, this node shifts the probability density to larger radii in
the confining potential and raises the energy, while $|N_4\rangle$ receives no
corresponding upward shift.
The resulting higher energy of the $|N_2\rangle$ configuration relative to
$|N_4\rangle$ provides a qualitative explanation for the ordering
$m^0_{N_2}>m^0_{N_4}$ found in our scattering fit.

We also examine the pole positions across all scenarios.
The Riemann sheets are labeled by four signs $(\pm,\pm,\pm,\pm)$ in the channel
order $(\pi N,\eta N,K\Lambda,K\Sigma)$.
For each channel, $+$ denotes the physical branch of the channel momentum, while
$-$ denotes its continuation across the corresponding right-hand cut.
Across Scenarios II\textendash{}IV, a subthreshold pole around
$1.00-0.17\,\mathrm{i}~\mathrm{GeV}$ appears on the \texttt{-+++} sheet, similar
to that reported in Ref.~\cite{Cao:2022zhn}.
On the \texttt{--++} and \texttt{---+} sheets, the poles in Scenarios III(a),
III(b), and IV(A)\textendash{}IV(C) correspond to the $N^*(1535)$ and
$N^*(1650)$ resonances, respectively.
While Scenarios IV(A, B) place the $N^*(1535)$ pole at a somewhat lower mass
($\sim1.500-0.075\,\mathrm{i}~\mathrm{GeV}$), Scenario IV(C) gives pole
positions at $1.529-0.055\,\mathrm{i}~\mathrm{GeV}$, in close agreement with the
PDG values~\cite{ParticleDataGroup:2026mpi}.
For the $N^*(1650)$, all three scenarios yield similar pole positions that are
consistent with the PDG values.
We also extract pole residues for both resonances in each channel.
After including phase-space factors and normalizing over the retained channels,
the resulting decay-width fractions are broadly comparable to PDG
compilations~\cite{ParticleDataGroup:2026mpi}.
For example, Scenario IV(C) gives $(0.45,0.55,0,0)$ for the $N^*(1535)$ in the
ordered channels $(\pi N,\eta N,K\Lambda,K\Sigma)$, and $(0.78,0.11,0.11,0)$ for
the $N^*(1650)$.
Furthermore, in Scenarios IV(B, C), pole-trajectory analysis connects the
$N^*(1535)$ primarily to the lighter bare state and the $N^*(1650)$ primarily to
the heavier one.
Thus, in Scenario IV(C), the $N^*(1535)$ is associated mainly with the
spin-$3/2$ bare configuration, whose coupling to $K\Lambda$ is suppressed, while
the $N^*(1650)$ is associated mainly with the spin-$1/2$ bare state and couples
more strongly to $K\Lambda$.
This pattern is consistent with the phenomenology summarized in
Ref.~\cite{ParticleDataGroup:2026mpi}, but is exactly reversed in Scenario
IV(B).
In Scenario IV(A), the mixing mechanism obscures this correspondence.

\begin{figure}[htbp]
  \centering
  \includegraphics[trim=0.72bp 0.72bp 0.72bp 0.72bp,clip,
    width=0.95\columnwidth]{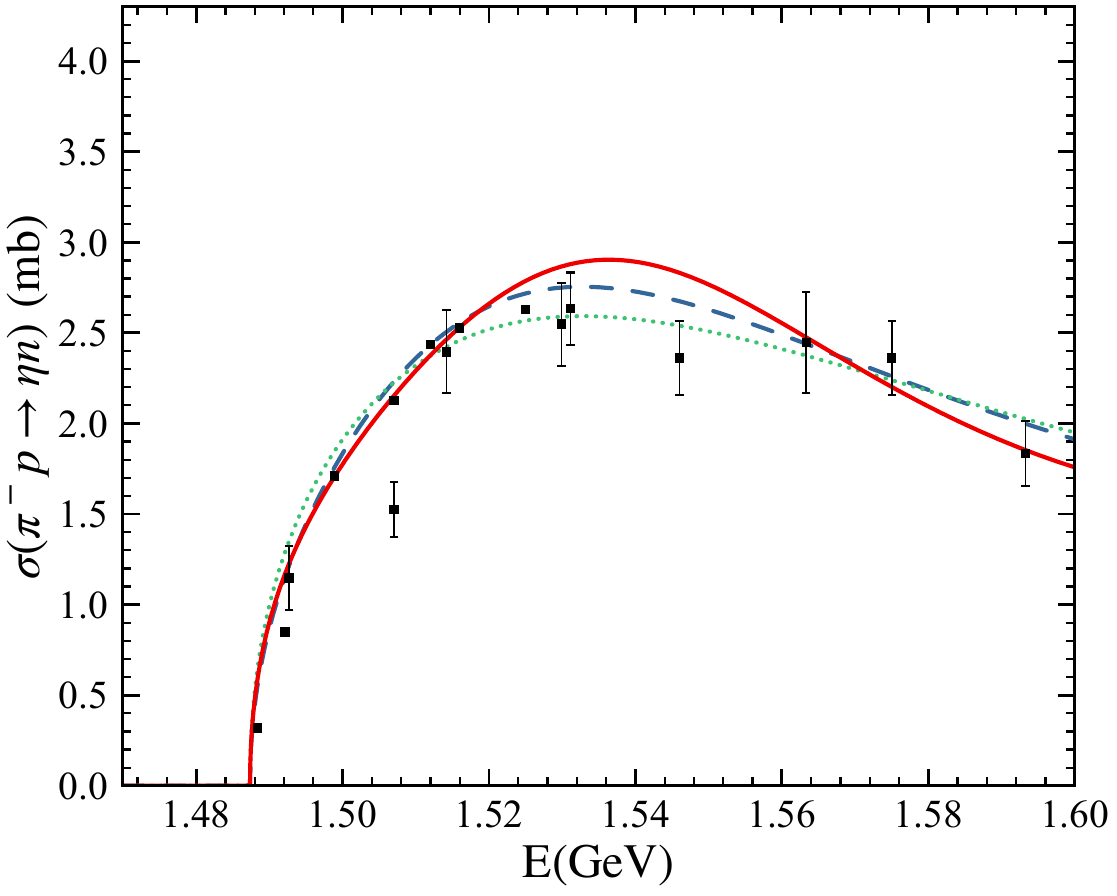}
  \caption{Total cross section $\sigma(\pi^- p \to \eta n)$ scaled by a factor
  of $1/1.2$ to account for three-body $\pi\pi N$ effects~\cite{Cieply:2013sya}.
  The blue dashed, green dotted, and red solid curves show Scenarios~IV(A),
  IV(B), and IV(C), respectively. The experimental data (black squares) are
  taken from the compilation in Ref.~\cite{Bruns:2019fwi}, originally measured
  in Refs.~\cite{Prakhov:2005qb, Bulos:1969ofe, Richards:1970cy,
  Feltesse:1975nz}.}
  \label{fig:sigma_etan}
\end{figure}

To further test the off-diagonal coupled-channel transition amplitudes
constrained by SU(3) flavor and quark-spin symmetries, we calculate the total
inelastic cross section $\sigma(\pi^- p \to \eta n)$ near the $N^*(1535)$ region
using the parameters fixed solely from elastic $\pi N$ scattering.
In the $s$-wave approximation below $1.60~\mathrm{GeV}$, the cross section is
related to the $I=1/2$ transition $T$ matrix by
\begin{equation}
  \label{eq:sigma_etan}
  \begin{aligned}
  \sigma(\pi^- p \to \eta n)\, ={}& \frac{2}{3} \frac{4\pi^3}{\left(k_{\pi
  N}^{\mathrm{on}}\right)^2} \rho_{\pi N}(E)\, \rho_{\eta N}(E) \\
  &\times\left| T_{\pi N, \eta N}(k_{\pi N}^{\mathrm{on}}, k_{\eta
  N}^{\mathrm{on}}; E) \right|^2 ,
  \end{aligned}
\end{equation}
where the factor $2/3$ comes from the isospin Clebsch\textendash{}Gordan
coefficients.
Because our two-body coupled-channel basis omits the three-body $\pi\pi N$
channel, two-body unitarity between the $\eta N$ and $K\Lambda$ thresholds
($1.487 < E < 1.611~\mathrm{GeV}$) routes the entire inelastic flux $1 -
\eta_{\pi N}^2$ into the $\eta N$ final state.
As noted by Ciepl\'y  and Smejkal~\cite{Cieply:2013sya, Bruns:2019fwi}, the
total reaction cross section implied by the SAID $\pi N$ inelasticity near
$\sqrt{s} = 1540~\mathrm{MeV}$ is approximately $20\%$ larger than the measured
$\pi^- p \to \eta n$ cross section peak due to the missing $\pi\pi N$ decay
flux.
Following Ref.~\cite{Cieply:2013sya}, we account for this three-body flux
dilution by dividing the calculated two-body cross section by an empirical
factor of $1.2$.

Figure~\ref{fig:sigma_etan} compares the scaled cross sections $\sigma(\pi^- p
\to \eta n)/1.2$ for Scenarios~IV(A), IV(B), and IV(C) with the experimental
data compiled in Ref.~\cite{Bruns:2019fwi} from Refs.~\cite{Prakhov:2005qb,
Bulos:1969ofe, Richards:1970cy, Feltesse:1975nz}.
Without introducing any additional free parameters, all three scenarios
reproduce the sharp threshold onset at $E\approx1.487~\mathrm{GeV}$ and the
$N^*(1535)$ peak profile up to $1.60~\mathrm{GeV}$, supporting the
symmetry-constrained $\pi N\to\eta N$ transition vertices.
Because all three scenarios describe $\eta_{\pi N}$ well in this energy window,
their predicted $\pi^- p\to\eta n$ cross sections remain closely clustered.
Among the three, Scenario IV(C) provides the best overall fit quality and pole
positions; its main departure from the conventional quark-model picture is the
inverted bare-mass ordering $m^0_{N_2}>m^0_{N_4}$, wherein the state with larger
total quark spin has the smaller bare mass.

In summary, current $\pi N$ data hardly distinguish among the three models and
only slightly favor Scenario IV(C).
Nevertheless, because infinite-volume scattering observables at the physical
point remain closely clustered among
Scenarios IV(A), IV(B), and IV(C), we carry all three solutions forward to the
finite-volume analysis in the next section to test them against lattice QCD
spectra.

\section{Finite-volume spectra}

The $\pi N$ $S_{11}$ data disfavor several scenarios, but the subcases of
Scenario IV give comparable fits.
We use finite-volume lattice QCD spectra as additional constraints to
distinguish these solutions.

At the physical pion mass, a complete set of coupled-channel amplitudes would
determine the finite-volume spectrum through L\"uscher's formalism.
In practice, the available data constrain mainly amplitudes with a $\pi N$
initial state. They do not determine the complete transition matrix among the
$\pi N$, $\eta N$, $K\Lambda$, and $K\Sigma$ channels.
This input does not uniquely fix the model parameters, which leads to the
degeneracy among the fits.
Finite-volume lattice spectra provide complementary constraints on the
coupled-channel dynamics.

The lattice spectra used here have two main limitations.
First, they are obtained at unphysical pion masses, so the quark-mass dependence
of the model must be specified.
These results also allow us to study hadron properties over a range of pion
masses.
Second, the calculations provide discrete levels at only a few box sizes.
For a multichannel system, the number of amplitude parameters can exceed the
number of available finite-volume constraints at a given energy. A
model-independent determination of the scattering matrix is then not feasible,
even for short-range interactions.
A predefined interaction ansatz is therefore mandatory, which makes the NPHF a
natural tool for this task.
The interactions shift the finite-volume levels away from the noninteracting
meson-baryon energies.
The pattern of these shifts, together with the eigenvector composition of each
state, provides information on the internal structure of the odd-parity nucleon
resonances.
We apply the formalism of Sec.~II to Scenario III(a) and Scenarios IV(A), IV(B),
and IV(C).

The lattice ensembles span a range of pion masses, so we include the required
dependence on $m_\pi^2$.
At leading order, only the bare masses are taken to depend explicitly on
$m_\pi^2$. The remaining interaction parameters are fixed in the spirit of
chiral perturbation theory, as proved in the $\rho$ meson
case~\cite{Wang:2025hew, Yu:2023xxf}.
The meson and baryon masses in each channel also vary with the pion mass.
We use the linear forms
\begin{align}
  m_M^2(m_\pi^2)&=m_M^2|_{\mathrm{phys}}+\alpha_M
  (m_\pi^2-m_\pi^2|_{\mathrm{phys}})\, ,\label{mesonextrapolation}\\
  m_B(m_\pi^2)&=m_B|_{\mathrm{phys}}+\alpha_B
  (m_\pi^2-m_\pi^2|_{\mathrm{phys}})\,
  \label{baryonextrapolation},
\end{align}
where the slopes $\alpha_M$ and $\alpha_B$ are determined from the PACS-CS
hadron masses~\cite{PACS-CS:2008bkb}.

For each bare nucleon state, we use the analogous form
\begin{equation}
  \label{barenucleonextrapolation}
  m^0_{N_i}(m_\pi^2)=m^0_{N_i}|_{\mathrm{phys}}+\alpha_{N_i}
  (m_\pi^2-m_\pi^2|_{\mathrm{phys}})\, ,
\end{equation}
where $\alpha_{N_i}$ is fitted to lattice QCD spectra. The $L\sim3$~fm ensembles
cover pion masses from 169 to 623~MeV in the Sommer scale-setting scheme.
In this scheme, the spatial extent varies with $m_\pi$.
We denote its extrapolated physical-point value by
$L|_{m_\pi=m_\pi^{\mathrm{phys}}}$.
The $m_\pi^2$ dependence of the box size is parametrized as
\begin{equation}
  \label{latticeextrapolation}
  L(m_\pi^2)=L{|_{{m_\pi } = m_\pi ^{\mathrm{phys}}}}+\beta
  (m_\pi^2-m_\pi^2|_{\mathrm{phys}})\, ,
\end{equation}
where $\beta$ is obtained by a linear interpolation of the lattice volumes.
The $L\sim2$~fm spectra provide additional constraints on the bare-state slopes.
For the $L\sim3$~fm ensembles, the physical-point extent is $2.99~\mathrm{fm}$,
increasing to $3.27~\mathrm{fm}$ at the heaviest pion mass.
For the $L\sim2$~fm ensembles, the corresponding extents are $1.95$ and
$2.12~\mathrm{fm}$.
In each scenario, the physical-point Hamiltonian parameters remain fixed by the
scattering fit. Only the bare-state slopes $\alpha_{N_i}$ are adjusted to the
finite-volume spectra.

We present two types of finite-volume results.
The first shows the energy levels as functions of $m_\pi^2$. Colored markers
identify the levels with the largest and second-largest overlap with each bare
state.
Within the assumed correspondence between local three-quark interpolators and
the bare basis, these markers indicate the levels expected to have stronger
overlap with such interpolators.
This criterion assumes that local three-quark interpolators couple
preferentially to the bare-state sector, so such states should preferentially be
observed with these operators. It gives a more selective test than comparing
energies alone.
No analogous bare-overlap criterion is imposed on levels extracted with explicit
meson-baryon interpolators.

The second type shows the bare-state components $|\langle B_j|E_i\rangle|^2$ and
channel-summed components $\sum_{\ell}|\langle\alpha(k_\ell)|E_i\rangle|^2$,
where $\ell$ runs over the retained momentum shells. Here, $\ket{E_i}$ is a
finite-volume eigenstate, while $\ket{B_j}$ and $\ket{\alpha(k_\ell)}$ belong to
the noninteracting model basis.
These quantities describe the composition of $\ket{E_i}$ within the NPHF basis.
This helps us to understand the internal structure of the
eigenstate in the finite volume.
As the interaction is switched on, the eigenstate with the largest bare-state
component can migrate between neighboring energy levels, but it remains
localized in the energy region associated with the corresponding bare mass.
The pion-mass dependence of these components therefore provides a stringent test
of the structural scenarios introduced in the previous section.

For $L\sim3$~fm, we use results from the CSSM
Collaboration~\cite{Mahbub:2013ala} and the Cyprus
Collaboration~\cite{Alexandrou:2014mka}. These spectra were obtained
predominantly with three-quark interpolators.
The exception is the lowest CSSM point at $m_\pi^2=0.08~\mathrm{GeV}^2$, which
was extracted with a five-quark interpolator and is expected to have a sizable
meson-baryon component~\cite{Kiratidis:2015vpa}.
For $L\sim2$~fm, we use the results of Lang and Verduci~\cite{Lang:2012db} and
the Hadron Spectrum Collaboration (HSC)~\cite{Edwards:2011jj}.
Because the correlation matrix of Lang and Verduci did not fully resolve the
excited states, we use only their lowest level. It was obtained with a nonlocal,
momentum-projected pion-nucleon interpolator.
The same bare-state slopes are used at both volumes. Their fitted values are
given separately for each scenario.

\begin{figure}[!tp]
   \centering
\includegraphics[width=1.0\columnwidth]{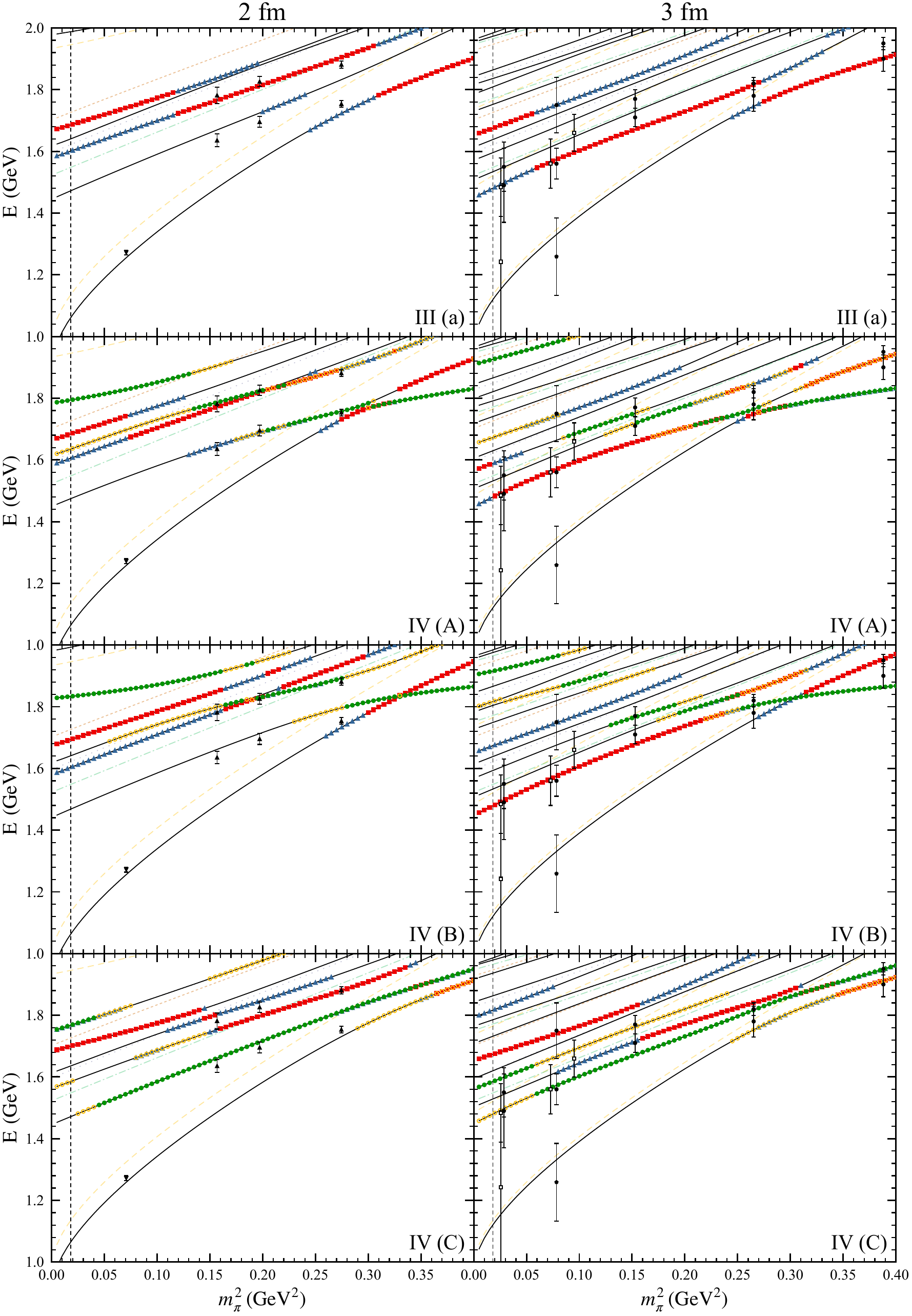}
\caption{Finite-volume spectra at $L\sim2~\mathrm{fm}$ (left) and
$L\sim3~\mathrm{fm}$ (right). The rows show Scenario III(a) and Scenarios IV(A),
IV(B), and IV(C), from top to bottom.
At $L\sim2~\mathrm{fm}$, the inverted triangles and triangles denote the results
of Lang and Verduci~\cite{Lang:2012db} and HSC~\cite{Edwards:2011jj},
respectively. At $L\sim3~\mathrm{fm}$, the circles and squares denote the
CSSM~\cite{Mahbub:2013ala} and Cyprus~\cite{Alexandrou:2014mka} results; the
pentagon marks the CSSM level extracted with a five-quark operator.
The vertical dashed line marks the physical pion mass, while the remaining
dashed curves show the corresponding noninteracting $\pi N$, $\eta N$,
$K\Lambda$, and $K\Sigma$ energy levels.
The solid black curves are the interacting NPHF eigenenergies.
Red squares and blue triangles mark the eigenstates with the largest and
second-largest overlap with the first bare state, respectively.
For Scenarios IV(A)--IV(C), filled green and open yellow circles mark the
largest and second-largest overlaps with the second bare state, respectively.
}
\label{fig:scEi23fm}
\end{figure}

\subsection{Finite-volume spectrum for Scenario III(a)}

Although Scenario III(a) gives a poorer $\pi N$ fit than Scenario IV, we also
test its finite-volume spectra.
With the physical-point Hamiltonian parameters fixed by the scattering data, the
only additional input is the bare-state slope $\alpha_N$.
Fitting both volumes gives
\begin{equation}
  \alpha_{N}=(0.869\pm0.076)~\mathrm{GeV}^{-1}.
\end{equation}

The first row of Fig.~\ref{fig:scEi23fm} shows the resulting spectra together
with the noninteracting meson-baryon energies.
The interactions produce visible energy shifts relative to the non-interacting
spectrum, and the lattice QCD levels can be associated with NPHF eigenstates in
this scenario.
Because the finite-volume spectrum is dense, nearly every lattice datum lies
close to a calculated eigenenergy.
Energy agreement alone is therefore not a selective test.
The lattice levels extracted with three-quark interpolators do not, however, lie
near the lower bands marked by large bare-state overlaps, shown as colored
markers in Fig.~\ref{fig:scEi23fm}.
This behavior is inconsistent with the interpolator--bare-state correspondence
assumed in our analysis.
We therefore find that Scenario III(a) does not give a satisfactory combined
description.

\subsection{Finite-volume spectrum for Scenario IV(A)}

Scenario~IV(A) retains both quark-model configurations and fixes their mixing
parameter to $\xi=-0.53$, with the bare-mass eigenvalues ordered as
$m_{B_1}<m_{B_2}$.
The additional bare state improves the description of the infinite-volume $\pi
N$ amplitude relative to Scenario~III(a).
The finite-volume spectra provide an independent constraint on the
two-bare-state dynamics.

With the physical-point Hamiltonian parameters held fixed, only the bare-state
slopes are adjusted to the lattice spectra.
A simultaneous fit at $L\sim2$ and $3~\mathrm{fm}$ gives
\begin{equation}
  \begin{aligned}
  \alpha_{N_2}&=(0.963\pm0.099)~\mathrm{GeV}^{-1}, \\
  \alpha_{N_4}&=(0.286\pm0.069)~\mathrm{GeV}^{-1}.
  \end{aligned}
\end{equation}
Because the second bare-mass eigenvalue is already heavy at the physical point
($1.864~\mathrm{GeV}$), the fit requires a small slope $\alpha_{N_4}$ to
approach the lattice levels at intermediate pion masses.

The second row of Fig.~\ref{fig:scEi23fm} shows the spectra for Scenario~IV(A).
Filled green and open yellow circles mark the largest and second-largest
overlaps with the second bare state.
At both $L\sim2$ and $3~\mathrm{fm}$, nearly all lattice levels at low and
intermediate pion masses lie near curves marked by large or second-largest
bare-state overlaps, which is a clear improvement over Scenario~III(a) and
favors the two-bare-state scenario.
However, because $\alpha_{N_4}$ is much smaller than $\alpha_{N_2}$, the level
ordering of the two bare states reverses around $m_\pi^2 \approx
0.27~\mathrm{GeV}^2$: the green-marked branch crosses below the red-marked
branch and flattens out at large pion masses.
Consequently, at $L\sim3~\mathrm{fm}$, the green-marked branch lies noticeably
below the lower lattice datum at the heaviest pion mass
($m_\pi^2\approx0.39~\mathrm{GeV}^2$), while the red- and yellow-marked branches
cluster near the upper datum.
In this case, the two lattice energy levels at the largest pion mass for
$L\sim3~\mathrm{fm}$ do not fall on the lowest energy level (i.e., the green
points), whereas in previous studies using the same three-quark operators, the
energy levels on the green branch could be clearly observed.

\subsection{Finite-volume spectrum for Scenario IV(B)}

Scenario~IV(B) sets the explicit mixing parameter to zero ($\xi=0$) while
retaining the conventional bare-mass ordering $m^0_{N_2}<m^0_{N_4}$.
With all physical-point parameters fixed by the scattering fit, the
finite-volume spectra give
\begin{equation}
  \begin{aligned}
  \alpha_{N_2}&=(1.159\pm0.10)~\mathrm{GeV}^{-1}, \\
  \alpha_{N_4}&=(0.177\pm0.072)~\mathrm{GeV}^{-1}.
  \end{aligned}
\end{equation}
Here the fitted slope $\alpha_{N_4}$ is even flatter than in Scenario~IV(A).%

The third row of Fig.~\ref{fig:scEi23fm} shows the levels for Scenario~IV(B).
Their overall distribution is qualitatively similar to that of Scenario~IV(A),
with the shallow green-marked branch again crossing below the red-marked branch
near $m_\pi^2\approx0.30~\mathrm{GeV}^2$ and underestimating the lower lattice
level at the heaviest pion mass at $L\sim3~\mathrm{fm}$.
These crossings indicate a dramatic reversal in the composition of the
corresponding energy levels, as shown in Figs.~\ref{fig:sc41contriEi3fm} and
\ref{fig:sc43contriEi3fm} in Appendix~\ref{app:component_figs}, where the states
$\ket{E_1}$ and $\ket{E_2}$ interchange their dominant components around
$m_\pi^2\approx0.27$--$0.32~\mathrm{GeV}^2$ as the meson-baryon thresholds
overtake the nearly flat $\ket{N_4}$ bare mass.

Compared with Scenario IV(A), at $L\sim3~\mathrm{fm}$ and the heaviest pion
masses, the lattice energy levels can reach/touch the lowest green-colored
level, a result that improves upon Scenario IV(A).
However, some of the energy levels still deviate from the color-coded curves.
In particular, for $L=2~\mathrm{fm}$, the lowest HSC data points at
$m_\pi^2=0.156~\mathrm{GeV}^2$ and $0.197~\mathrm{GeV}^2$ lie on an unmarked
curve with negligible bare-state overlap.
Thus, neither Scenario~IV(A) nor IV(B) describes the lattice spectra across the
full pion-mass perfectly.

\subsection{Finite-volume spectrum for Scenario IV(C)}

\begin{figure}[tbp]
    \centering
    \includegraphics[width=0.48\textwidth]{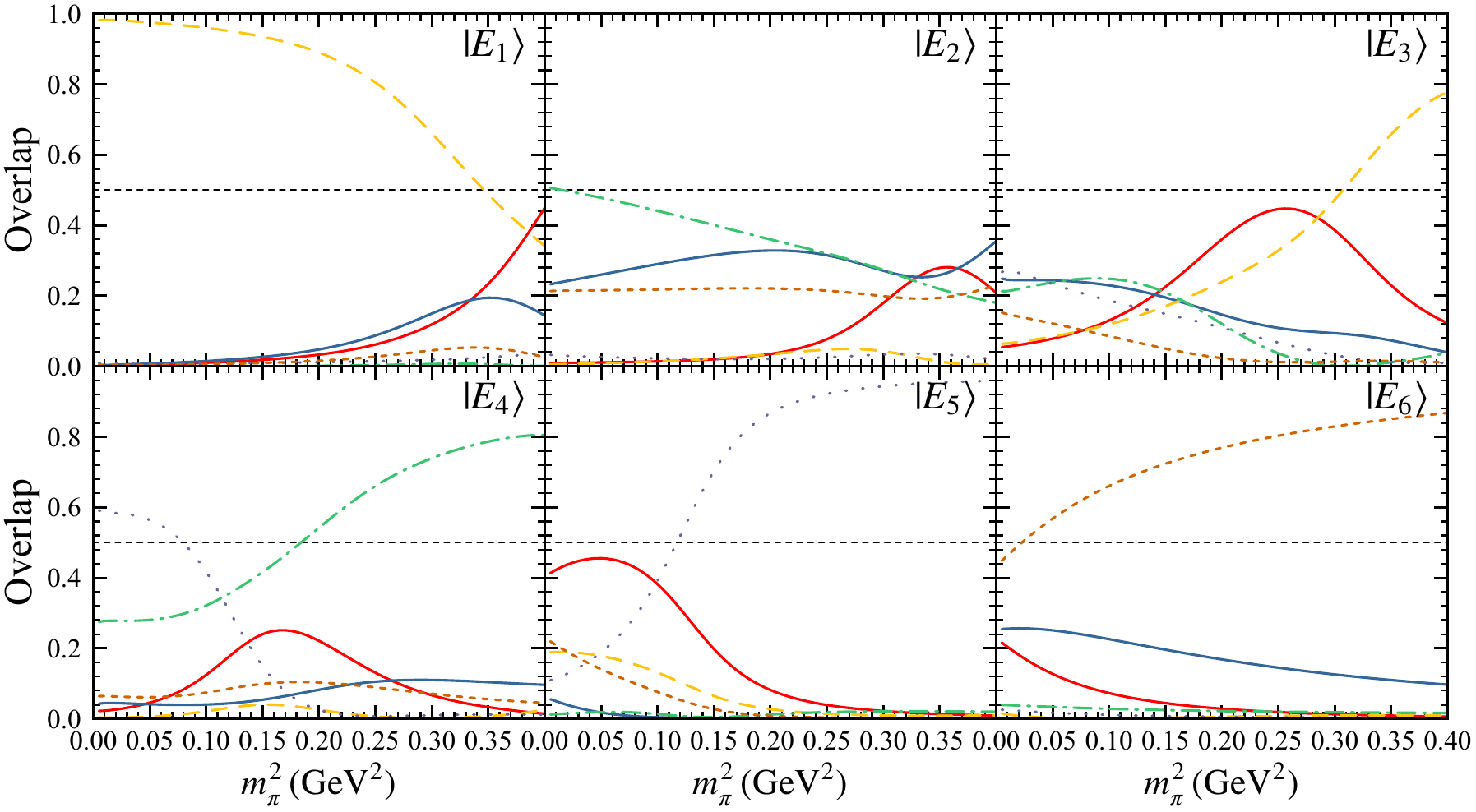}
    \par\medskip
    \includegraphics[width=0.48\textwidth]{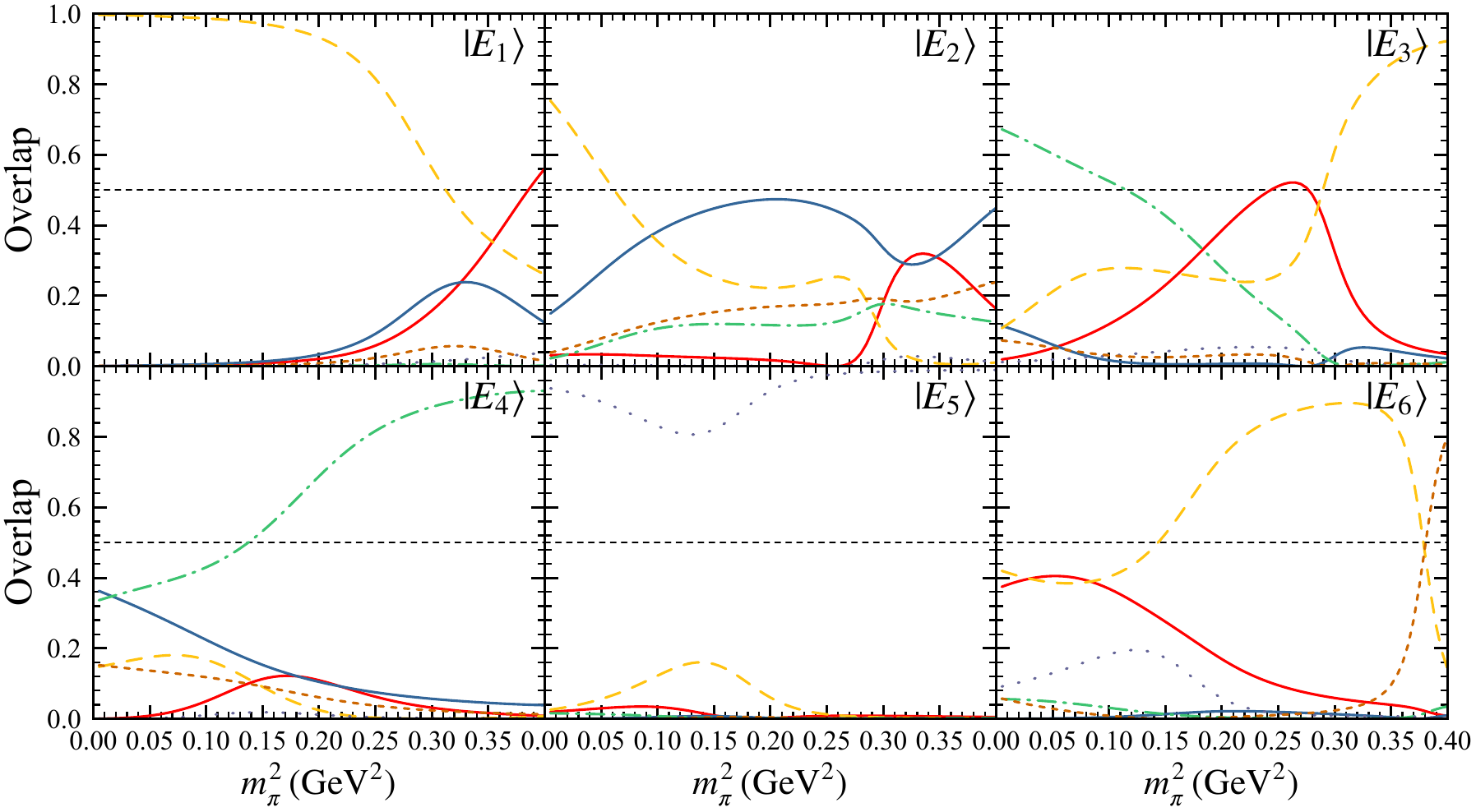}
    \caption{Model-basis decompositions of the six lowest finite-volume
    eigenstates in Scenario IV(C), shown as $|\langle B_j|E_i\rangle|^2$ and
    $\sum_{\ell}|\langle\alpha(k_\ell)|E_i\rangle|^2$. The upper and lower six
    panels correspond to $L\sim2$ and $3~\mathrm{fm}$, respectively. The solid
    curves show the $N_2$ (red) and $N_4$ (blue) components. The remaining
    curves show the $\pi N$ (yellow dashed), $\eta N$ (green dash-dotted),
    $K\Lambda$ (purple dotted), and $K\Sigma$ (brown short-dashed) components.}
    \label{fig:sc44contriEi2fm}
    \label{fig:sc44contriEi3fm}
\end{figure}

Scenario~IV(C) also sets the explicit mixing parameter to zero ($\xi=0$) and
features the reversed bare-mass ordering $m^0_{N_2}>m^0_{N_4}$ at the physical
point.
The finite-volume fit adds only the two bare-state slopes,
\begin{equation}
  \begin{aligned}
  \alpha_{N_2}&=(0.619\pm0.091)~\mathrm{GeV}^{-1}, \\
  \alpha_{N_4}&=(1.168\pm0.10)~\mathrm{GeV}^{-1}.
  \end{aligned}
\end{equation}
Unlike Scenarios~IV(A) and IV(B), both bare states now exhibit substantial
positive quark-mass slopes.

The fourth row of Fig.~\ref{fig:scEi23fm} shows the spectra for Scenario~IV(C),
which improve noticeably upon those of Scenarios~IV(A) and IV(B).
The fit results of Scenario~IV(C) demonstrate a globally superior performance
compared to the two preceding models, Scenarios~IV(A) and IV(B). This
improvement is primarily manifested in two aspects as follows.

First, at both $L=2~\mathrm{fm}$ and $L=3~\mathrm{fm}$, all lattice energy
levels fall squarely on the color-coded model curves across the entire pion-mass
range.
In particular, at $L\sim2~\mathrm{fm}$ across all three pion masses (including
the two HSC points at $m_\pi^2=0.274~\mathrm{GeV}^2$), and at
$L\sim3~\mathrm{fm}$ for the first four ensembles, both the lower and upper HSC
levels lie near eigenstates carrying the largest bare-state overlaps, in full
agreement with the assumed correspondence between three-quark interpolators and
the bare basis.

Second, because both bare-mass slopes are sizable and positive, no level-order
reversal occurs at large pion masses.
Instead, the two bare-state branches rise together and undergo a smooth crossing
near $m_\pi^2\simeq0.30~\mathrm{GeV}^2$.
Consequently, at $L\sim3~\mathrm{fm}$ and the heaviest pion mass
($m_\pi^2\approx0.39~\mathrm{GeV}^2$), the two highest lattice levels precisely
occupy the two lowest energy branches, showing excellent agreement with the
model values.
This stands in clear contrast to Scenarios~IV(A) and IV(B), where a reversal or
flattening of the branches misplaced these levels.

Fig.~\ref{fig:sc44contriEi3fm} shows the model-basis decomposition of the six
lowest eigenstates. The corresponding plots for the other scenarios are given in
Appendix~\ref{app:component_figs}.
Near the physical pion mass, the lowest level $\ket{E_1}$ at both volumes is
dominated by the $\pi N$ channel.
The lower excited levels $\ket{E_2}$ and $\ket{E_3}$ (together with $\ket{E_4}$
at $L\sim3~\mathrm{fm}$) develop sizable $N_4$ components mixed with the $\pi N$
and $\eta N$ channels, which the infinite-volume pole analysis links mainly to
the $N^*(1535)$.
The fourth level $\ket{E_4}$ is largely dominated by the $\eta N$ channel, while
the $K\Lambda$ channel dominates $\ket{E_5}$ at $L\sim3~\mathrm{fm}$.
Substantial $N_2$ components appear in the higher levels\textemdash{}primarily
$\ket{E_5}$ and $\ket{E_6}$ at $L\sim2~\mathrm{fm}$ and $\ket{E_6}$ at
$L\sim3~\mathrm{fm}$\textemdash{}which the pole analysis links mainly to the
$N^*(1650)$.
As $m_\pi^2$ increases toward $0.30~\mathrm{GeV}^2$, the bare-state components
migrate into the lowest three eigenstates $\ket{E_1}$--$\ket{E_3}$ and mix
strongly near the bare-mass crossing, with the $N_2$ component overtaking $N_4$
in $\ket{E_1}$ for $m_\pi^2\gtrsim0.30~\mathrm{GeV}^2$ while $N_4$ becomes
dominant in $\ket{E_2}$ above $m_\pi^2\approx0.35~\mathrm{GeV}^2$.
Even for levels with the largest bare-state overlaps, no single bare-state
component generally exceeds $50\%$.
Meson-baryon components are therefore essential within this description of the
excited nucleons.

Overall, Scenario~IV(C) gives the best combined description of the
infinite-volume scattering observables and the finite-volume spectra at both
volumes among the scenarios considered.

\section{Scattering-state contamination}
\label{sec:scattering_contamination}

\subsection{Contamination-function formalism}
\label{sec:scattering_contaminationformalism}

Within the NPHF, Hamiltonian eigenvectors provide a model-based estimate of
scattering-state contamination in lattice QCD correlators constructed from local
three-quark interpolators.
An interacting finite-volume eigenstate can couple to such an interpolator
through its bare-state component even when it also contains sizable meson-baryon
components. These additional levels can affect the extraction of
single-particle-like energies.
The contamination function measures the residual correlator strength after the
eigenstates with the largest bare-state overlaps are removed.
The single- and two-bare-state constructions are derived in Sec.~VII of
Ref.~\cite{Abell:2023nex}. Here, we apply the two-state construction to Scenario
IV(C).

The main equations of contamination functions are as
follows~\cite{Abell:2023nex},
\begin{align}
G_j(t)&=\sum_n\left|\alpha_j\langle N_2|E_n\rangle+\beta_j\langle
N_4|E_n\rangle\right|^2\, e^{-E_nt},\label{eq:gj}\\
C_j(t)&=\frac{1}{G_j(t)}\sum_{\substack{n\notin\{n_2^{(1,2)}, \\ n_4^{(1,2)}\}}}
\left|\alpha_j\langle N_2|E_n\rangle +\beta_j\langle
N_4|E_n\rangle\right|^2e^{-E_nt}.\label{eq:cj}
\end{align}
where $n_a^{(1)}$ and $n_a^{(2)}$ label the levels with the largest and
second-largest $|\langle N_a|E_n\rangle|^2$ for each bare state $N_a$, and
\begin{equation}\label{eq:alphabeta}
  \begin{split}
    \alpha_j&=\langle N_2|E_{n_j^{(1)}}\rangle,\\
    \beta_j&=\langle N_4|E_{n_j^{(1)}}\rangle.
  \end{split}
\end{equation}
Here, $\ket{E_{n_j^{(i)}}}$ state is the largest($i=1$) and
second-largest($i=2$) $j$-th bare-state components.
In the NPHF, the overlaps between bare states and eigenstates are calculated
from the eigenvectors of the Hamiltonian.
Furthermore, if we have two operators satisfying
$\bar{\chi}_i\ket{\Omega}=\ket{N_i}$, we will obtain that,
\begin{equation}
  \label{eq:twobarestateoperator}
(\alpha^*\bar{\chi}_1+\beta^*\bar{\chi}_2)\, \ket{\Omega}
  =\alpha^*\ket{N_2}+\beta^*\ket{N_4}.
\end{equation}
Then these four factors $\alpha_{1,2}$ and $\beta_{1,2}$ can also be recognized
as $\alpha_j=\langle \Omega|\chi_1|E_{n_j^{(1)}}\rangle$ and $   \beta_j=\langle
\Omega|\chi_2|E_{n_j^{(1)}}\rangle$, which is determined by the lattice inputs.

We consider two local three-quark interpolators for the odd-parity nucleon
channel:
$\chi_1=\epsilon^{abc}(u^{aT}C\gamma_5d^b)\, u^c$ and
$\chi_2=\epsilon^{abc}(u^{aT}Cd^b)\, \gamma_5u^c$.
In the spin assignment used here, the scalar-diquark structure of $\chi_1$ gives
no direct overlap with the total three-quark spin-$3/2$ configuration. We
therefore associate that configuration primarily with $\chi_2$.
The spin-$1/2$ bare state can also overlap with $\chi_2$, but we assume that its
overlap with $\chi_1$ is dominant.
This assignment is compatible with the relativistic wave-function analysis
discussed in Sec.~\ref{sec:infinite_volume_fits} ~\cite{Mickley:2026zsi}.

Using this bare-subspace assignment, we compare $C_j(t)$ using coefficients from
the lattice correlation matrix and from the NPHF Hamiltonian eigenvectors.

\begin{figure*}[!tp]
  \centering
  \includegraphics[width=\textwidth]{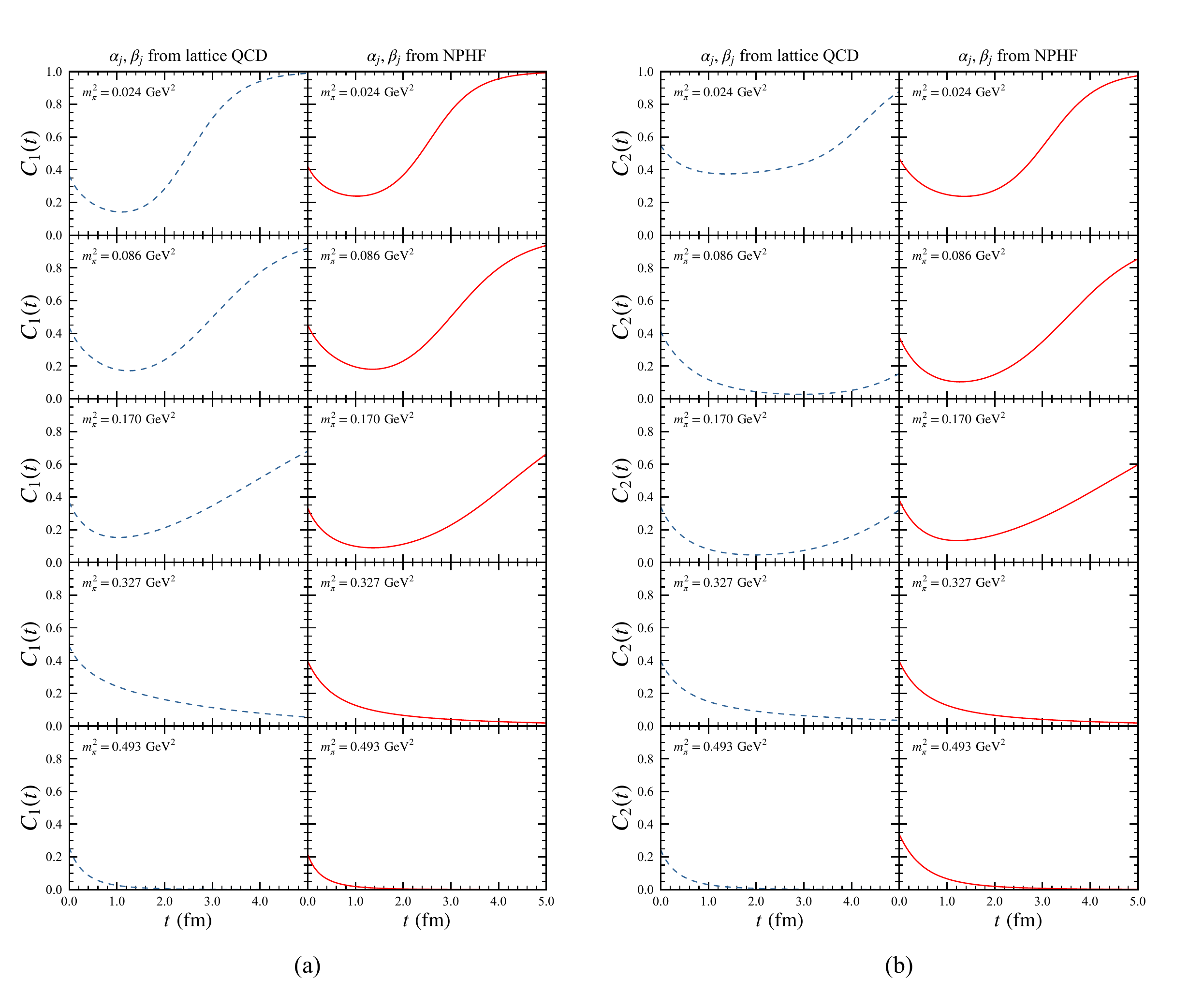}
  \caption{Contamination functions for Scenario IV(C) at $L\sim3~\mathrm{fm}$
  for the five pion masses used in Ref.~\cite{Mahbub:2013ala}.
  The pion mass increases from top to bottom.
  For each optimized correlator, the functions are evaluated using
  Eq.~\eqref{eq:cj}.
  The red solid curves use $\alpha_j$ and $\beta_j$ from the NPHF Hamiltonian
  eigenvectors, while the blue dashed curves use the lattice-QCD
  correlation-matrix eigenvectors.}
  \label{fig:sc44Cit3fm}
\end{figure*}

\subsection{Contamination functions for Scenario IV(C)}

\begin{figure}[htbp]
  \centering
  \includegraphics[width=\columnwidth]{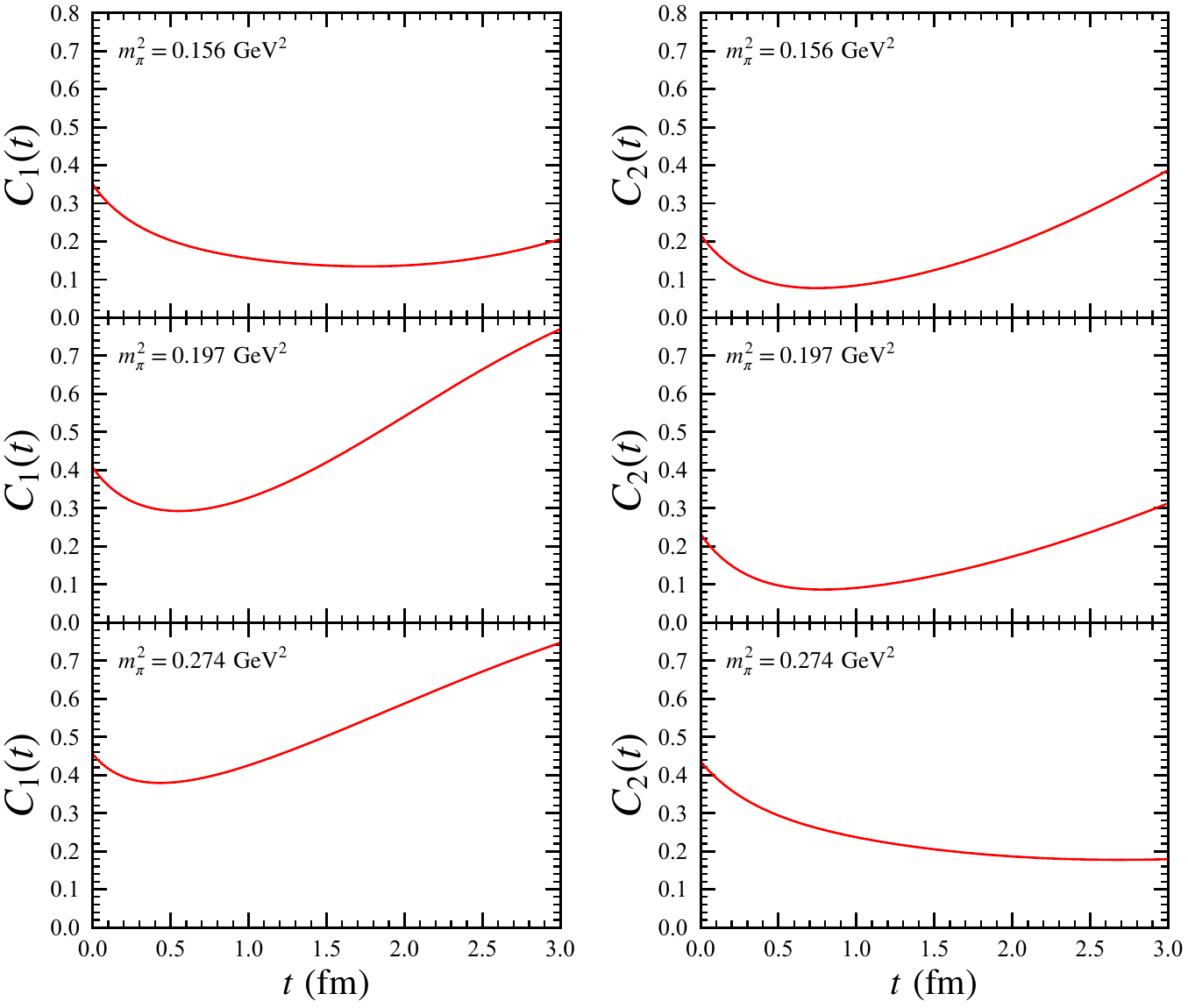}
  \caption{Contamination functions for Scenario IV(C) at $L\sim2~\mathrm{fm}$.
  The three rows correspond to the pion masses used in the HSC
  comparison~\cite{Edwards:2011jj}, increasing from top to bottom. Because
  lattice correlation-matrix eigenvectors are unavailable at this volume, all
  curves use coefficients from the NPHF Hamiltonian eigenvectors.}
  \label{fig:sc44Cit2fm}
\end{figure}

Figs.~\ref{fig:sc44Cit3fm} and \ref{fig:sc44Cit2fm} show $C_j(t)$ at $L\sim3$
and $2~\mathrm{fm}$, respectively. We use the interaction parameters and
bare-state slopes fixed in the preceding analyses, with no additional fit
parameters.

At $L\sim3$~fm, the coefficients entering Eq.~\eqref{eq:cj} can be determined
from the lattice-QCD correlation-matrix eigenvectors of
Ref.~\cite{Mahbub:2013ala}.
That calculation used an $8\times8$ correlation matrix built from the two
interpolators with four smearing levels at both source and sink.
We use the dominant 100-sweep components shown in Fig.~11(a) of
Ref.~\cite{Mahbub:2013ala}.
The coefficients $\alpha_1$ and $\alpha_2$ are taken from the $\chi_1$ ($u_5$ in
Ref.~\cite{Mahbub:2013ala}) components for states 1 and 2.
The coefficients $\beta_1$ and $\beta_2$ are taken from the corresponding
$\chi_2$ ($u_6$ in Ref.~\cite{Mahbub:2013ala}) components.
The relative signs of these components are retained because they determine the
interference term in the optimized overlap.
For comparison, we also determine the coefficients from the NPHF Hamiltonian
eigenvectors through Eq.~\eqref{eq:alphabeta}.

Depending on the pion mass and the optimized correlator, the functions shown in
Fig.~\ref{fig:sc44Cit3fm} either decrease over the displayed Euclidean-time
interval or develop a minimum before increasing again.
This behavior reflects the location of the removed bare-dominant states relative
to the lower-lying states that remain in the spectral sum.
A comparison of the two coefficient sets tests whether the NPHF eigenvectors
reproduce the operator-mixing trends in the lattice correlation matrix.
For $C_1(t)$, associated with $\chi_1$ and predominantly with $\ket{N_2}$, the
two determinations have broadly similar trends.
This is consistent with the predominantly spin-$1/2$ assignment of $\chi_1$ used
above.
The differences are more visible for $C_2(t)$.
For the three lowest pion masses, both the lattice-based and NPHF curves rise at
longer Euclidean times, but the NPHF curves rising more rapidly.
This difference is consistent with $\chi_2$ having an additional overlap with
the spin-$1/2$ configuration that is not captured by a one-to-one operator
mapping.
Across the five pion masses, the magnitudes, minima, and detailed time
dependence remain sensitive to the coefficient source, but the overall trends
are broadly similar.
Within the assumed operator mapping, this comparison supports the use of the
NPHF eigenvectors as a model estimate of the contamination. This further
corroborates that the $|N_2\rangle$ bare state is heavier than the $|N_4\rangle$
bare state, a conclusion that is supported by the lattice data.

Fig.~\ref{fig:sc44Cit2fm} shows the results at $L\sim2~\mathrm{fm}$.
Only the NPHF-derived coefficients are available at this volume.
Their qualitative behavior is compatible with the $L\sim3~\mathrm{fm}$ results.
We await independent verification from future lattice-QCD calculations.

\section{Conclusion}

We have studied the odd-parity nucleon resonances $N^*(1535)$ and $N^*(1650)$ in
the $S_{11}$ partial wave within the NPHF.
The coupled-channel model contains two bare states, $|N_2\rangle$ and
$|N_4\rangle$, with total quark spins $1/2$ and $3/2$, respectively. It also
includes four meson-baryon channels.
Under the imposed SU(3) flavor and spin constraints, the scenarios considered
here contain at most 11 fit parameters. These constraints also impose the
vanishing $|N_4\rangle$--$K\Lambda$ coupling.
The infinite-volume $\pi N$ data alone leave several model solutions. We
therefore use finite-volume lattice QCD spectra as an independent constraint.
We also use model-basis overlaps and the contamination function to assess
scattering-state contributions to lattice correlators.
Together, the scattering fits, finite-volume spectra, and eigenvector analysis
distinguish four classes of scenarios.

Our main conclusions are summarized as follows.
First, the model containing both three-quark bare states and the four channels
$\pi N$, $\eta N$, $K\Lambda$, and $K\Sigma$ describes both the infinite-volume
amplitude and the finite-volume spectra. It performs better than the
single-bare-state, pure coupled-channel, and pure bare-state models considered
here.
Scenario~IV(C), which features unmixed bare states with the inverted mass
ordering $m^0_{N_2}>m^0_{N_4}$, gives the most consistent description among the
scenarios considered.
It reproduces the $\pi N$ $S_{11}$ phase shift and inelasticity and describes
the large-pion-mass lattice levels that are missed by the other scenarios.
The lattice levels also lie near NPHF eigenstates with appreciable bare-state
overlaps, consistent with their extraction using predominantly three-quark
interpolators.

Second, we have examined the case of normal bare-state mass ordering.
However, we found that neither the fit to the infinite-volume T-matrix nor the
interpretation of finite-volume energy levels is as appropriate as the inverse
mass ordering model.
Therefore, this work favors the inverse mass ordering model, i.e., the total
three-quark spin-$1/2$ state $|N_2\rangle$ is heavier than the spin-$3/2$ state
$|N_4\rangle$.
The combined scattering and finite-volume analyses favor this ordering within
the model.
The contamination-function analysis is consistent with the associated operator
assignment.
The result is also consistent with a recent study of relativistic quark wave
functions.
That study relates the ordering to the interplay between a node in the lower
Dirac component and the confining potential~\cite{Mickley:2026zsi}.
We believe that more theoretical and experimental studies are certainly needed
in the future.

Third, no single bare-state component generally exceeds $50\%$, even in levels
with the largest bare-state overlaps. These levels contain substantial
meson-baryon components.
Within the NPHF description, coupled-channel effects are therefore essential for
both resonances. The $N^*(1535)$ and $N^*(1650)$ arise from the interplay
between the bare configurations and the meson-baryon continuum.

Several extensions would test this picture.
Lattice correlation-matrix eigenvectors at smaller volumes would provide an
independent test of the operator mapping and contamination function. Higher
statistics and a larger interpolator basis would strengthen this comparison.
Higher-order chiral interactions and additional channels would test the model
dependence of the results.
The microscopic origin of the bare-mass inversion also requires further
quark-model and lattice QCD studies.
The same combination of scattering fits, lattice spectra, and contamination
diagnostics can be applied to other baryon resonances.

\begin{acknowledgments}
The authors thank Feng-Kun Guo, Bing-Song Zou and Qiang Zhao for the useful
discussion.
Especially, Professor Zhao Qiang suggested that we carefully consider the zero
interaction between the total quark spin-3/2 bare state and $\Lambda K$, which
prompted us to build this model to re-study the two nucleon states.
This work is supported by the National Natural Science Foundation of China under
Grants No.~12221005,
by the Chinese Academy of Sciences under Grant No. YSBR-101.
It was also supported by the University of Adelaide and by the Australian
Research Council through Discovery Project DP230101791 (AWT).

\end{acknowledgments}

\appendix

\section{Robustness of the bare-mass ordering at lower bare-mass scales}
\label{app:lower_bound}

\begin{figure*}[htbp]
  \centering
 \includegraphics[width=0.98\textwidth]{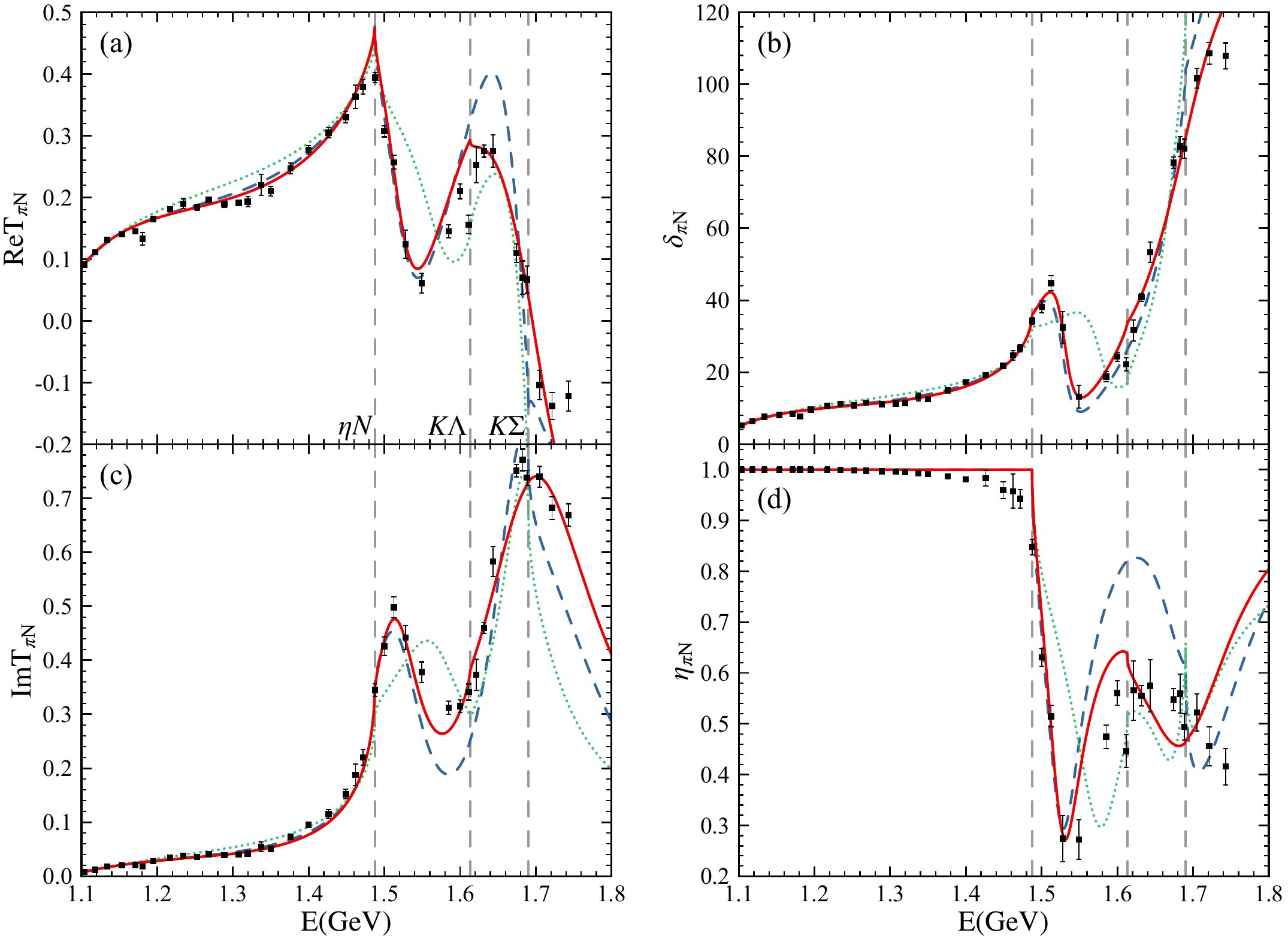}
  \caption{
    Real part of the $T$ matrix (a), phase shift (b), imaginary part of the $T$
    matrix (c), and inelasticity (d) for Scenarios~IV(A) (blue dashed), IV(B)
    (green dotted), and IV(C) (red solid) when the lighter bare mass is
    constrained to $\le 1.600~\mathrm{GeV}$ (Table~\ref{tab:fit_params_app}).
    The remaining notation follows Fig.~\ref{fig:sc4}.
  }
  \label{fig:T_IV_app}
\end{figure*}

In the main text (Table~\ref{tab:scenarios-combined}), coupled-channel
self-energy dressing places the lighter bare state at $1.750~\mathrm{GeV}$,
where Scenarios~IV(A), IV(B), and IV(C) yield comparable infinite-volume fits
($\chi^2/\mathrm{dof}=6.133$, $6.785$, and $5.491$, respectively) and require
finite-volume lattice QCD spectra to lift their degeneracy.
Because constituent quark models typically place the lowest $P$-wave nucleon
excitations around $1.50$--$1.65~\mathrm{GeV}$~\cite{Isgur:1978xj,
Capstick:1986ter, Bijker:2000gq, Zhao:2006an}, it is instructive to test how the
three scenarios behave when the lighter bare state ($m_{B_1}$ in IV(A),
$m_{N_2}^0$ in IV(B), and $m_{N_4}^0$ in IV(C)) is constrained to lie at or
below $1.600~\mathrm{GeV}$.

As shown in Table~\ref{tab:fit_params_app} and Fig.~\ref{fig:T_IV_app}, lowering
the bare-mass scale directly breaks the infinite-volume degeneracy and strongly
favors Scenario~IV(C).
Scenario~IV(C) retains an accurate description of all four scattering
observables across the entire energy region ($\chi^2/\mathrm{dof}=6.908$), with
stable resonance poles at $1.516-0.044\,\mathrm{i}~\mathrm{GeV}$ and
$1.688-0.094\,\mathrm{i}~\mathrm{GeV}$ and a clear inverted splitting between
$m_{N_2}^0=1.862~\mathrm{GeV}$ and $m_{N_4}^0=1.600~\mathrm{GeV}$ ($\Delta
m=0.262~\mathrm{GeV}$).
By contrast, the conventional mass-ordering scenarios degrade severely:
Scenario~IV(A) ($\chi^2/\mathrm{dof}=17.555$) substantially overestimates
$\mathrm{Re}\,T_{\pi N}$ and $\eta_{\pi N}$ between $1.60$ and
$1.68~\mathrm{GeV}$, while Scenario~IV(B) ($\chi^2/\mathrm{dof}=29.487$) fails
to reproduce the resonance structures across $1.50$--$1.70~\mathrm{GeV}$ and
shifts the $N^*(1535)$ pole upward to $1.575-0.062\,\mathrm{i}~\mathrm{GeV}$.
Moreover, because two-channel unitarity between the $\eta N$ and $K\Lambda$
thresholds relates $\sigma(\pi^- p \to \eta n)$ directly to $1-\eta_{\pi N}^2$,
the distorted inelasticity in Scenario~IV(B) also leads to large discrepancies
with the $\pi^- p \to \eta n$ cross section, whereas Scenario~IV(C) preserves
good agreement.

This sharp contrast reflects the symmetry-constrained coupling patterns in
Table~\ref{tab:Gcoupling}.
When the lighter bare state is pushed down to
$1.600~\mathrm{GeV}$\textemdash{}directly into the resonance
region\textemdash{}the scattering amplitude becomes highly sensitive to its
channel couplings.
Placing the spin-$3/2$ state $\ket{N_4}$ (which has a weaker bare $\pi N$
coupling and a vanishing $K\Lambda$ coupling) at $1.600~\mathrm{GeV}$ and the
strongly coupled spin-$1/2$ state $\ket{N_2}$ at $1.862~\mathrm{GeV}$ allows
Scenario~IV(C) to naturally maintain the destructive interference dip between
the $N^*(1535)$ and $N^*(1650)$.
Thus, imposing a lower quark-model-like bare-mass scale already rules out
Scenarios~IV(A) and IV(B) at the infinite-volume level, further reinforcing the
inverted bare-mass hierarchy $m_{N_2}^0 > m_{N_4}^0$ established in the main
text.

\begin{table}[htbp]
  \caption{
    Fit parameters, $\chi^2/\mathrm{dof}$, and pole positions for
    Scenarios~IV(A), IV(B), and IV(C), when the lighter bare mass is constrained
    to $\le 1.600~\mathrm{GeV}$, for comparison with
    Table~\ref{tab:scenarios-combined}.
    Case~(A) uses the fixed quark-model mixing $\xi=-0.53$, where the
    asterisk-prefixed entries denote the bare-mass eigenvalues $m_{B_1}$ and
    $m_{B_2}$; cases~(B) and (C) are unmixed ($\xi=0$).
    The three pole rows correspond to the Riemann sheets \texttt{-+++},
    \texttt{--++}, and \texttt{---+}, following the conventions of
    Table~\ref{tab:scenarios-combined}.
    Underlined entries lie at a fit boundary and are shown without errors.
  }
  \label{tab:fit_params_app}
  \begingroup
  \setlength{\tabcolsep}{3pt}
  \renewcommand{\arraystretch}{1.08}
  \small
  \begin{ruledtabular}
  \begin{tabular}{lccc}
    Parameter & IV(A) & IV(B) & IV(C) \\
    \hline
    $g_0$ & $0.722\pm0.0050$ & $0.450\pm0.0053$ & $0.535\pm0.0069$ \\
    $\Lambda_{\pi N}/\mathrm{GeV}$ & $\underline{0.400}$ & $\underline{0.400}$ &
    $0.432\pm0.0036$ \\
    $\Lambda_{\eta N}/\mathrm{GeV}$ & $\underline{0.400}$ & $\underline{1.600}$
    & $0.460\pm0.019$ \\
    $\Lambda_{K\Lambda}/\mathrm{GeV}$ & $\underline{0.400}$ &
    $\underline{1.600}$ & $\underline{1.600}$ \\
    $\Lambda_{K\Sigma}/\mathrm{GeV}$ & $0.402\pm0.023$ & $\underline{0.400}$ &
    $\underline{0.400}$ \\
    $m^0_{N_2}/\mathrm{GeV}$ & $*\,\underline{1.600}$ & $\underline{1.600}$ &
    $1.862\pm0.0045$ \\
    $m^0_{N_4}/\mathrm{GeV}$ & $*\,1.738\pm0.0016$ & $\underline{1.900}$ &
    $\underline{1.600}$ \\
    $g_{N_2}$ & $1.055\pm0.015$ & $1.516\pm0.027$ & $0.572\pm0.0066$ \\
    $g_{N_4}$ & $1.156\pm0.032$ & $1.043\pm0.015$ & $1.409\pm0.026$ \\
    $\Lambda_{N_2}/\mathrm{GeV}$ & $0.757\pm0.0087$ & $0.512\pm0.0057$ &
    $\underline{1.600}$ \\
    $\Lambda_{N_4}/\mathrm{GeV}$ & $0.715\pm0.015$ & $1.179\pm0.016$ &
    $0.694\pm0.010$ \\
    \hline
    $\chi^2/\mathrm{dof}$ & 17.555 & 29.487 & 6.908 \\
    \hline
    \multirow{3}{*}{Pole$/\mathrm{GeV}$} & $1.000-0.158\,\mathrm{i}$ &
    $0.993-0.178\,\mathrm{i}$ & $0.993-0.171\,\mathrm{i}$ \\
    \cline{2-4}
     & $1.518-0.050\,\mathrm{i}$ & $1.575-0.062\,\mathrm{i}$ &
     $1.516-0.044\,\mathrm{i}$ \\
    \cline{2-4}
     & $1.682-0.056\,\mathrm{i}$ & $1.694-0.040\,\mathrm{i}$ &
     $1.688-0.094\,\mathrm{i}$ \\
  \end{tabular}
  \end{ruledtabular}
  \endgroup
\end{table}

\section{Model-basis decompositions for the remaining scenarios}
\label{app:component_figs}

For completeness, we show the model-basis decompositions of the six lowest
finite-volume eigenstates for Scenario III(a) and Scenarios IV(A) and IV(B).
The corresponding results for Scenario IV(C) appear in
Fig.~\ref{fig:sc44contriEi3fm}.
For each eigenstate, the bare-state components are shown separately, while each
meson-baryon component is summed over the retained finite-volume momenta.
These normalized NPHF components characterize the eigenstates within the chosen
model basis and are not direct lattice QCD observables.

\begin{figure*}[!tbp]
    \centering
    \includegraphics[width=0.74\textwidth]{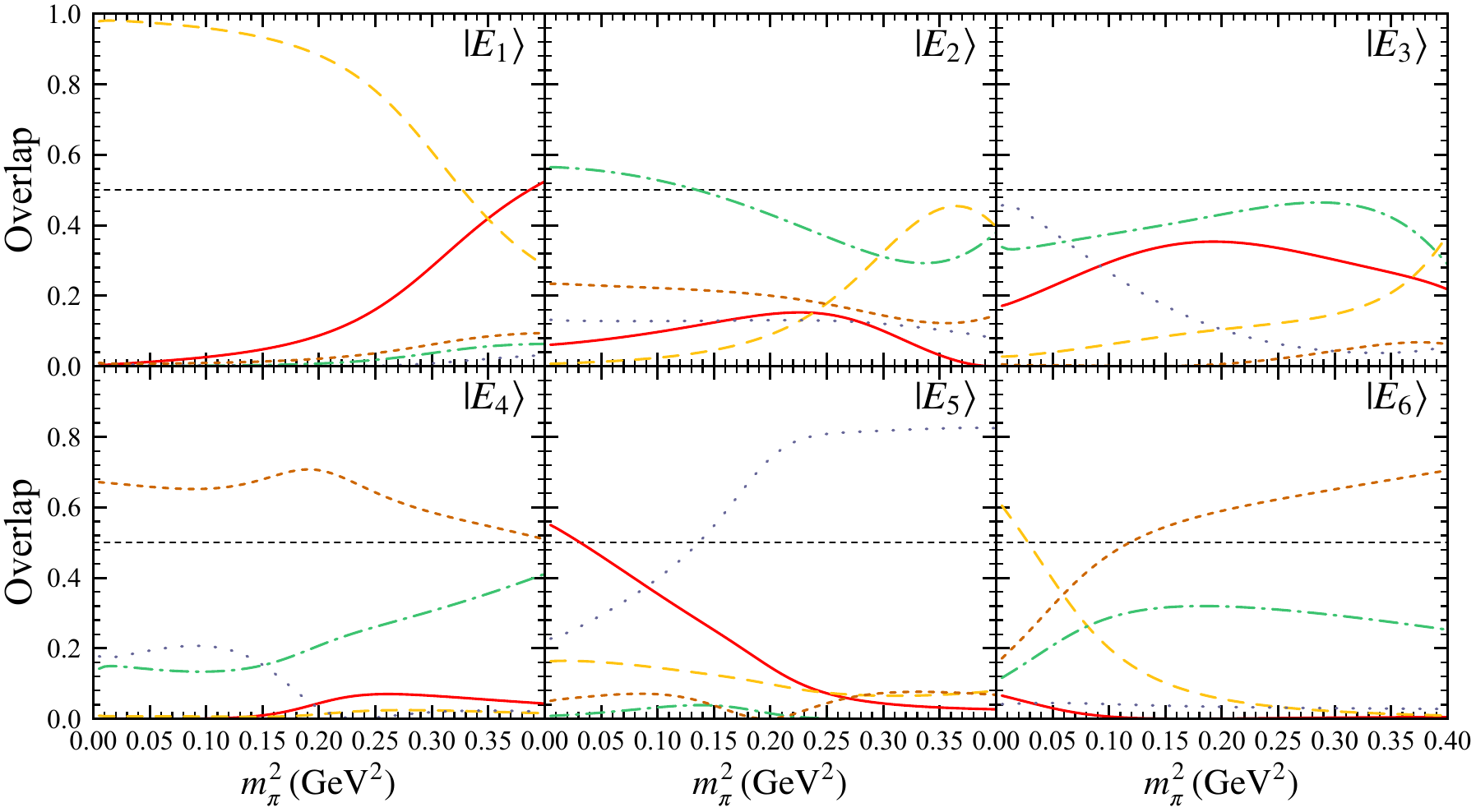}
    \includegraphics[width=0.74\textwidth]{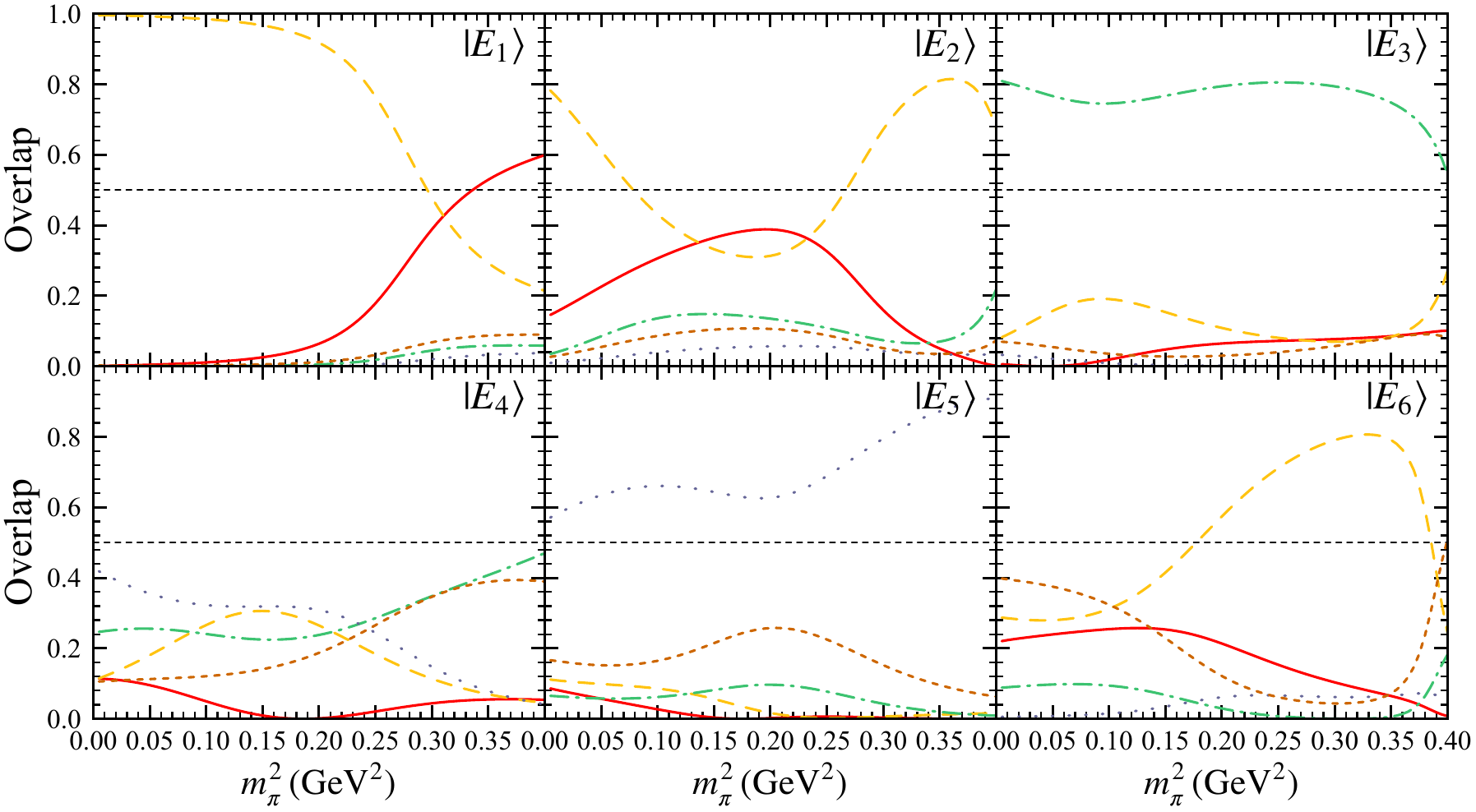}
    \caption{Model-basis decompositions of the six lowest finite-volume
    eigenstates in Scenario III(a), shown as $|\langle B_j|E_i\rangle|^2$ and
    $\sum_{\ell}|\langle\alpha(k_\ell)|E_i\rangle|^2$. The upper and lower six
    panels correspond to $L\sim2$ and $3~\mathrm{fm}$, respectively. The solid
    red curves show the bare-state component. The remaining curves show the $\pi
    N$ (yellow dashed), $\eta N$ (green dash-dotted), $K\Lambda$ (purple
    dotted), and $K\Sigma$ (brown short-dashed) components.}
    \label{fig:sc31contriEi3fm}
\end{figure*}

\begin{figure*}[!tbp]
    \centering
    \includegraphics[width=0.74\textwidth]{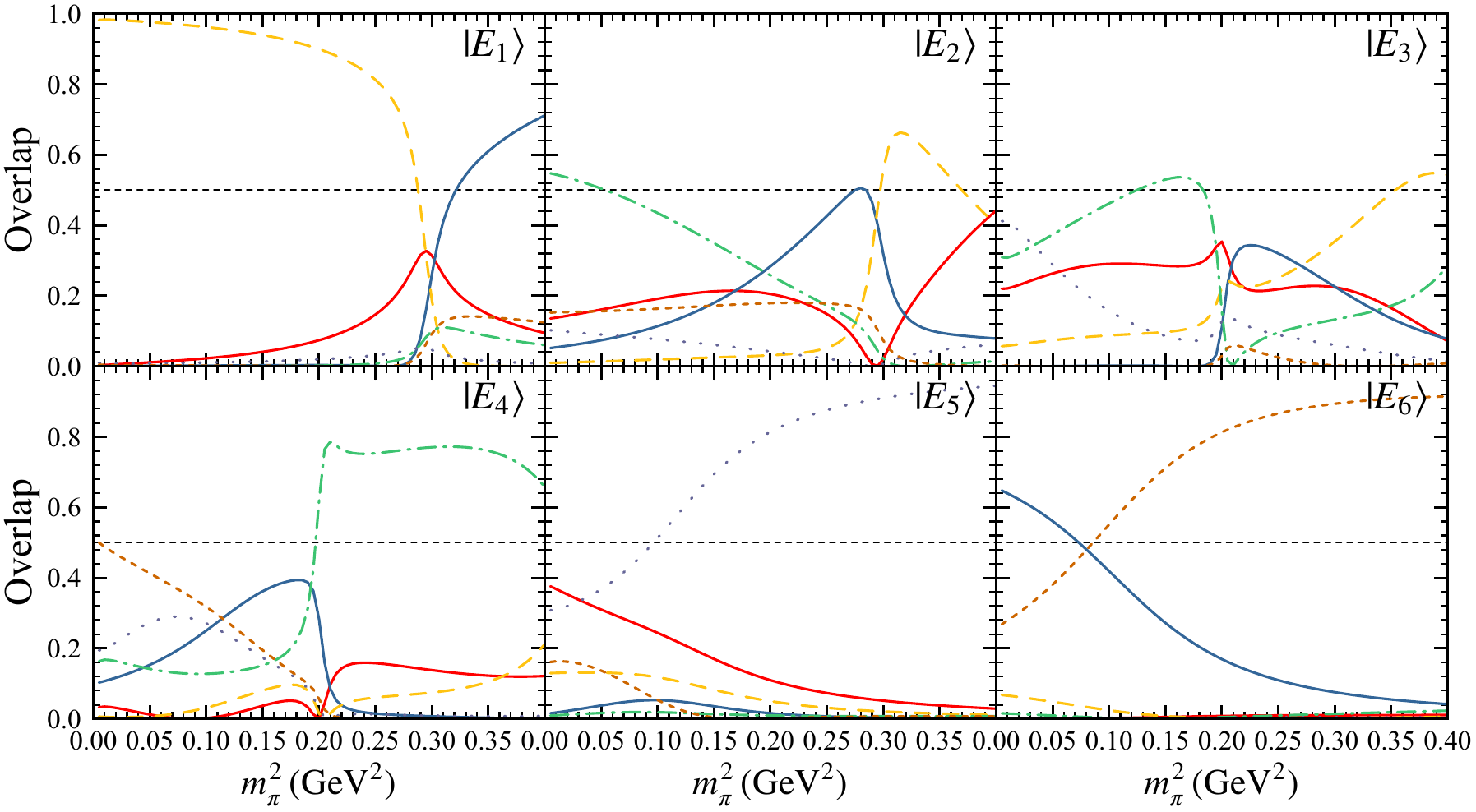}
    \par\medskip
    \includegraphics[width=0.74\textwidth]{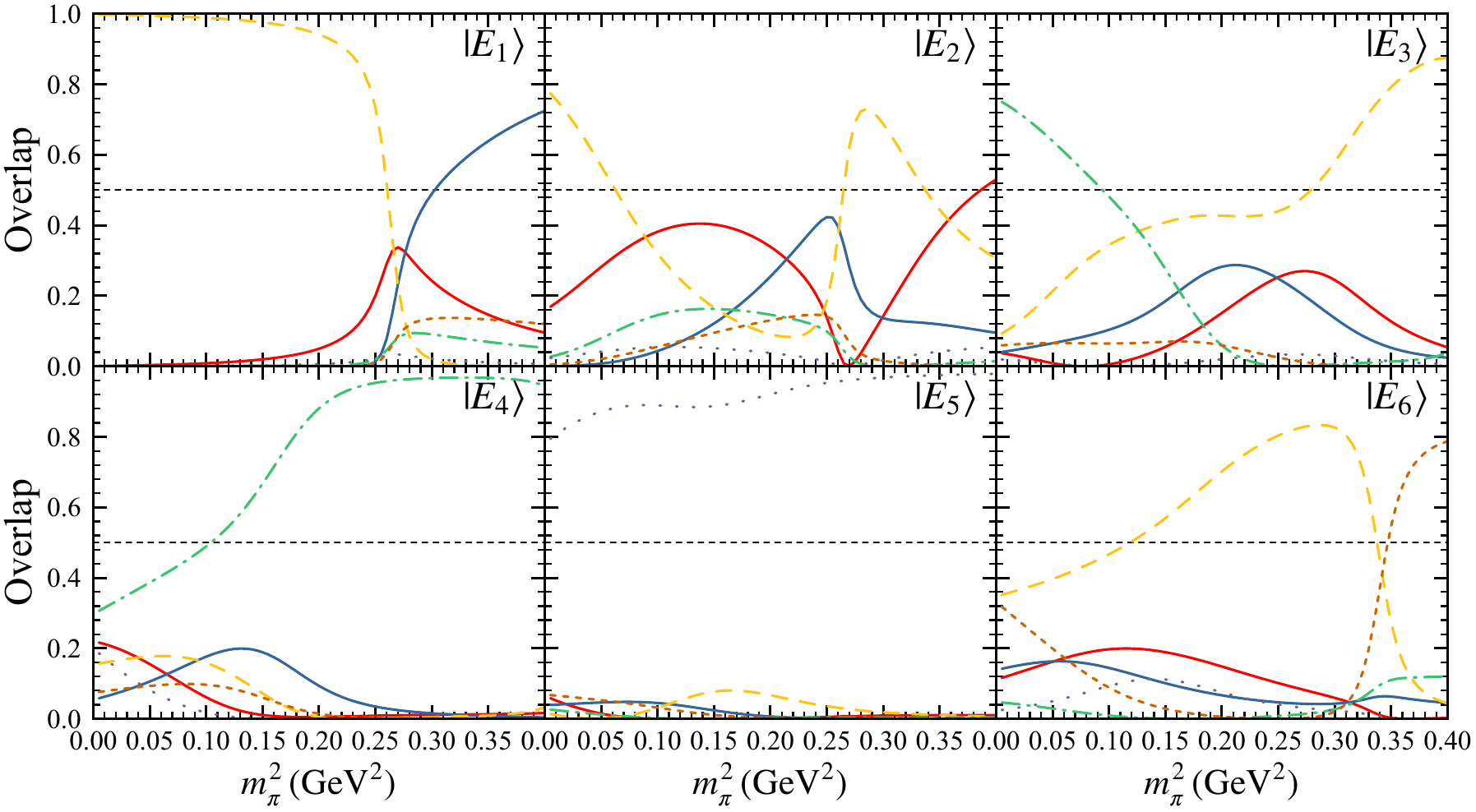}
    \caption{Model-basis decompositions of the six lowest finite-volume
    eigenstates in Scenario IV(A), shown as $|\langle B_j|E_i\rangle|^2$ and
    $\sum_{\ell}|\langle\alpha(k_\ell)|E_i\rangle|^2$. The upper and lower six
    panels correspond to $L\sim2$ and $3~\mathrm{fm}$, respectively. The solid
    curves show the $N_2$ (red) and $N_4$ (blue) components. The remaining
    curves show the $\pi N$ (yellow dashed), $\eta N$ (green dash-dotted),
    $K\Lambda$ (purple dotted), and $K\Sigma$ (brown short-dashed) components.}
    \label{fig:sc41contriEi2fm}
    \label{fig:sc41contriEi3fm}
\end{figure*}

\begin{figure*}[!tbp]
    \centering
    \includegraphics[width=0.74\textwidth]{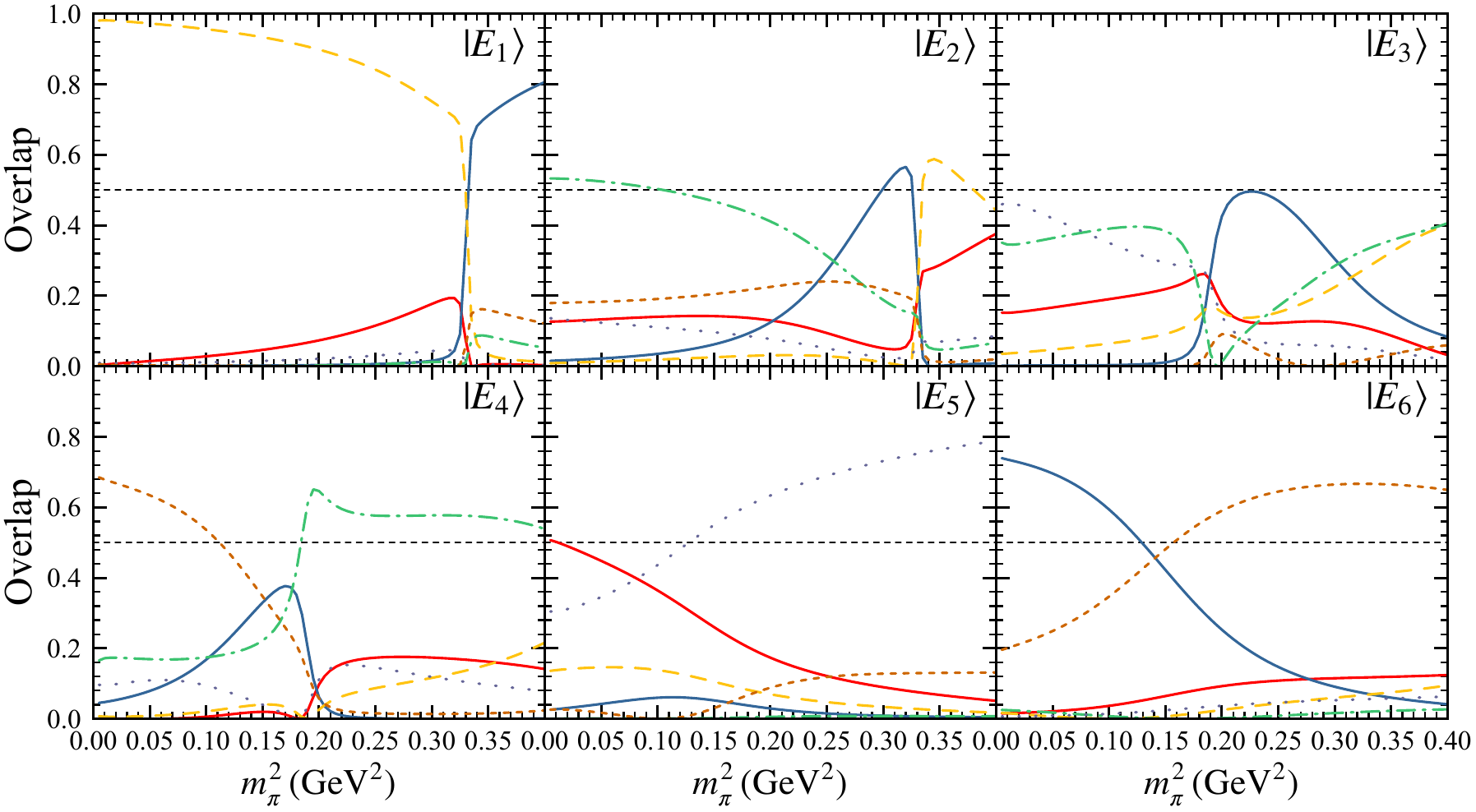}
    \par\medskip
    \includegraphics[width=0.74\textwidth]{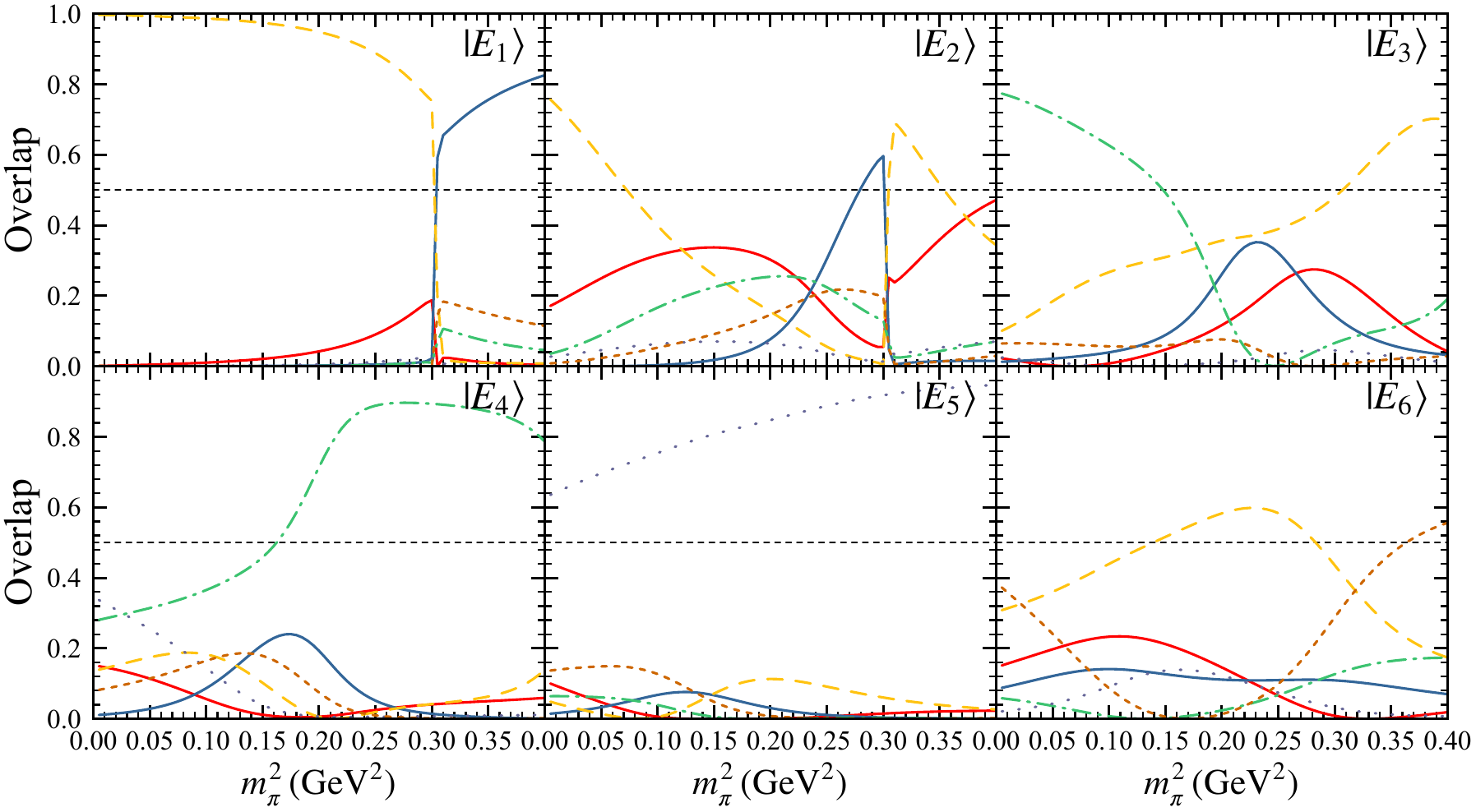}
    \caption{Model-basis decompositions of the six lowest finite-volume
    eigenstates in Scenario IV(B), shown as $|\langle B_j|E_i\rangle|^2$ and
    $\sum_{\ell}|\langle\alpha(k_\ell)|E_i\rangle|^2$. The upper and lower six
    panels correspond to $L\sim2$ and $3~\mathrm{fm}$, respectively. The solid
    curves show the $N_2$ (red) and $N_4$ (blue) components. The remaining
    curves show the $\pi N$ (yellow dashed), $\eta N$ (green dash-dotted),
    $K\Lambda$ (purple dotted), and $K\Sigma$ (brown short-dashed) components.}
    \label{fig:sc43contriEi2fm}
    \label{fig:sc43contriEi3fm}
\end{figure*}

\clearpage
\bibliography{references}

\end{document}